\documentclass[11pt]{article}
\usepackage[a4paper]{geometry}
\usepackage[utf8]{inputenc}
\usepackage{graphicx}
\usepackage{dsfont}
\usepackage{fancyhdr}
\usepackage{booktabs}
\usepackage[colorlinks=true, allcolors=blue,pagebackref=true]{hyperref}
\usepackage{url}
\usepackage{xcolor}
\usepackage{colortbl}
\usepackage{subfig}
\usepackage{mathtools}
\usepackage[T1]{fontenc}
\usepackage{comment}
\usepackage{natbib}
\usepackage{eurosym}
\usepackage{amsmath}
\allowdisplaybreaks
\usepackage{xcolor}
\usepackage{enumitem}
\usepackage{float}
\usepackage{here}
\usepackage{wrapfig}
\usepackage{lscape}
\usepackage{bbold}
\usepackage{stmaryrd}
\usepackage[export]{adjustbox}
\usepackage{tabularx}
\usepackage{amssymb}
\usepackage{mathtools}
\usepackage{multirow,makecell}
\usepackage[french,onelanguage, boxruled, vlined]{algorithm2e}
\usepackage{amsthm, bm}
\usepackage[french,onelanguage, boxruled, vlined]{algorithm2e}
\usepackage{lscape}
\usepackage{ulem}
\usepackage{vmargin}
\usepackage[french,onelanguage, boxruled, vlined]{algorithm2e}
\usepackage{booktabs}
\newcommand{\HRule}{\rule{\linewidth}{0.5mm}}

\newcommand{\dsum}{\displaystyle\sum}
\newcommand{\dprod}{\displaystyle\prod}

\usepackage{authblk}

\renewcommand*{\backref}[1]{}
\renewcommand*{\backrefalt}[4]{}

\setmarginsrb{2cm}{1cm}{2cm}{1cm}{0.5cm}{0.5cm}{0.5cm}{0.5cm}

\title{A Bayesian shared-frailty spatial scan statistic model for time-to-event data. }
\author[,1,3]{Camille Frévent\thanks{Corresponding author: \texttt{camille.frevent@univ-lille.fr}}}
\author[1]{Mohamed-Salem Ahmed}
\author[2,3]{Sophie Dabo-Niang}
\author[1]{Michaël Genin}

\affil[1]{Univ. Lille, CHU Lille, ULR 2694 - METRICS: Évaluation des technologies de santé et des pratiques médicales, F-59000 Lille, France.}

\affil[2]{Univ. Lille, CNRS, UMR 8524 - Laboratoire Paul Painlevé, F-59000 Lille, France.}

\affil[3]{MODAL team, INRIA Lille-Nord Europe, France.}

\date{}

\begin{document}

\maketitle

\noindent \HRule
\begin{center} \textbf{Abstract} \end{center} 
Spatial scan statistics are well known and widely used methods for the detection of spatial clusters of events. In the field of spatial analysis of time-to-event data, several models of scan statistics have been proposed. However, these models do not take into account the potential intra-unit spatial correlation of individuals nor a potential correlation between spatial units. To overcome this problem, we propose here a scan statistic based on a Cox model with shared frailty that takes into account the spatial correlation between spatial units. In simulation studies, we have shown that (i) classical models of spatial scan statistics for time-to-event data fail to maintain the type I error in the presence of intra-spatial unit correlation, and (ii) our model performs well in the presence of both intra-spatial unit correlation and inter-spatial unit correlation. Our method has been applied to epidemiological data and to the detection of spatial clusters of mortality in patients with end-stage renal disease in northern France.\\

\HRule \\

\section{Introduction}
In many applications, researchers seek to determine if there exist unusual spatial aggregations of data, namely spatial clusters. In the field of public health, epidemiologists seek to identify the presence, within a geographical area, of spatial clusters in which the risk of disease is unusually higher (or lower), thus making it possible to (i) formulate hypotheses to guide etiological research, and (ii) conduct localized public health policies. As another example, in the field of environmental sciences, researchers may be interested in determining the presence of environmental black spots defined by particularly unusual pollutant concentrations in a specific area, thus leading to local actions to prevent or solve the problem. 

Among the statistical methods for detecting spatial clusters, spatial scan statistics are widely used methods to detect statistically significant spatial clusters with a scanning window and without any pre-selection bias. These methods were introduced by \cite{spatialdisease} and \cite{kulldorff1997spatial} in the cases of Bernoulli and Poisson models. 
Since then, the approach has been extended to many other spatial data distributions. For example, in the univariate framework, Gaussian \citep{normalkulldorff}, ordinal \citep{jung2007spatial}, zero-inflated \citep{canccado2014spatial, de2015spatial,  canccado2017bayesian} and Poisson with overdispersion \citep{zhang2012spatial, de2015spatial} models were proposed. In the context of multivariate or functional data, several spatial scan statistics have also been developed  \citep{kulldorffmulti, neill2010multivariate, a_multivariate_gaussian,frevent2021detecting,frevent2021investigating,smida2022wilcoxon}. The reader is referred to \cite{abolhassani2021up} for a more complete review of spatial scan statistics. These methods have been widely used in various fields of application such as epidemiology \citep{green2006population, marciano2018epidemiological, genin2020fine, khan2021geographic}, environmental science \citep{wan2020industrial,shi2021spatial}, oncology \citep{leiser2020spatial}, criminology \citep{minamisava2009spatial} and astronomy \citep{de2008star}.

In the field of spatial epidemiology, the study of the spatial distribution of time-to-event data can identify areas in which the survival time of patients is different from the rest of the geographical area (e.g., the probability of survival is lower than elsewhere). From an epidemiological point of view, the identification of these areas of unusual survival time is particularly useful to search for local risk factors that condition the survival of patients. Moreover, this information can help public health decision-makers to conduct targeted and specific local policies.  In the context of spatial cluster detection of time-to-event data, \cite{huang2007spatial} and \cite{bhatt2014spatial} respectively proposed spatial scan statistics based on an exponential and a Weibull model. More recently, \cite{usman2018log} proposed a parametric model considering a log-Weibull distribution. Although these methods are widely used in practice to detect spatial clusters of time-to-event data \citep{gregorio2007place, henry2009geographic, wan2012access}, they are totally parametric. A first semi-parametric method using a Cox model has been proposed by \cite{cook2007spatial}.

Unlike other spatial scan statistics models, these models consider data measured at the individual level. However, in the field of health data, the exact geographic location of patients is rarely known (i.e., for reasons of anonymity), as patients are rather located through an administrative spatial unit (e.g., municipalities). In this context, the above mentioned methods are based on the strong assumption of independence between observations, which is a classical assumption in the field of spatial scan statistics. This assumption leads to two major drawbacks. First, these methods do not take into account the potential correlation between individuals belonging to the same spatial unit, namely intra-spatial unit correlation. The latter can be induced by unmeasured characteristics of the spatial units (e.g., health care supply) that affect the survival time of patients \citep{austin2017tutorial}. Second, these methods do not take into account potential spatial correlation between spatial units. However, by nature, spatial data can present a spatial correlation and it is expected that geographically close units are more correlated than distant ones \citep{li2009modeling}. Furthermore, it has been shown that ignoring spatial correlation when using spatial scan statistics leads to a significant increase in the type I error \citep{loh2007accounting}. Since then, several authors have proposed methods that take spatial correlation into account \citep{loh2007accounting, lin2014generalized, lee2020spatial,ahmed2021spatial}. However, none of these methods have been designed for time-to-event data and allow for intra-unit spatial correlation.

In the field of time-to-event data analysis, models have been proposed to take into account unobserved factors common to groups of individuals, e.g., members of the same family share a common genetics, patients in the same hospital receive similar care, etc. One way to take into account this intra-group homogeneity is to introduce a random effect common to all individuals in a group, namely shared frailty \citep{clayton1978model,liang1995some,Hougaard2000}. The shared frailties are assumed to be independent between groups \citep{liang1995some}. However, in the case where the groups correspond to spatial units, such an assumption is unrealistic since close spatial units tend to be correlated \citep{arlinghaus1995practical}. To this end, \cite{li2002modeling} proposed an extension of the shared frailty models to the case of spatially correlated frailty, allowing to take into account not only the intra-spatial unit correlation, but also the possible correlation between the spatial units.
This approach has been used in many application studies \citep{banerjee2003frailty, ojiambo2013modeling, aswi2020bayesian}. \\

\noindent This paper develops a new spatial scan statistic for time-to-event data based on a Bayesian semi-parametric Cox model with spatially correlated shared frailties. Section \ref{sec:method} describes the methodological aspects of the scan statistic model. Section \ref{sec:simulation} proposes both the design and the results of simulation studies that aim at evaluating (i) the performance of classical methods in the presence of intra-unit spatial correlation and (ii) the performance of our approach in the presence of both intra-unit spatial correlation and spatial correlation. Section \ref{sec:realdata} describes the application of our method to epidemiological data and the detection of spatial clusters of mortality in patients with end-stage renal disease in northern France. Lastly, the results are discussed in Section \ref{sec:discussion}.

\section{Methodology} \label{sec:method}

\subsection{General principle}
Let us consider $K$ non-overlapping spatial locations $s_1, \dots,s_k,\dots, s_K$ of an observation domain $S \subset \mathbb{R}^2$ and $i_1^{(k)}, \dots,i_n^{(k)},\dots, i_{N_k}^{(k)}$ be  $N_k$ individuals belonging to spatial location $s_k$. The total number of individuals in $S$ is defined by $N=\sum_{k=1}^KN_k$. Here we are interested in the time-to-event data measured on individuals: one note respectively $T_{i_n^{(k)}}$ and $\delta_{i_n^{(k)}}$ the observation time of the $i_n$th individual in spatial location $s_k$ and $\delta_{i_n^{(k)}}$ the censoring indicator which is equal to 0 if the individual $i_n^{(k)}$ is censored and 1. In the following, we only consider the cases of right censoring (i.e., the event of interest could not have occurred before the beginning of the study), censoring is assumed to be uninformative, and the event times are supposed to be independent from the censoring times. 

We seek to test the presence of spatial clusters in which individuals present shorter (or longer) survival times compared to other individuals observed in the rest of $S$. In this context, spatial scan statistics are designed to detect spatial clusters and to test their statistical significance by testing a null hypothesis $\mathcal{H}_0$ (the absence
of a cluster) against a composite alternative hypothesis $\mathcal{H}_1$ (the presence of at least one cluster
$w \subset S$ presenting abnormal time-to-event values). Following \cite{cressie}, a spatial scan statistic is the maximum of a concentration index over a set
of potential clusters $\mathcal{W}$. In the following and without loss of generality, we focus on variable-size
circular clusters. Hence in line with \cite{kulldorff1997spatial}, the set of potential circular clusters $\mathcal{W}$ can be defined by:
$\mathcal{W} = \{w_{k,l} / 1 \le |w_{k,l}| \le \frac{N}{2}, 1 \le k, l \le K\}$, where $w_{k,l}$ is the disc centred on $s_k$ that passes through $s_l$ and $|w_{k,l}|$ is the number of individuals in $w_{k,l}$: a cluster comprises at most 50\% of the studied population (i.e., $N/2$) \citep{spatialdisease}.
Remark that in the literature other possibilities have been proposed such as elliptical clusters \citep{elliptic}, rectangular clusters \citep{rectangular} or arbitrarily shaped clusters \citep{tango2005flexibly,zhou2015adaptive,yin2018hybrid}.

\subsection{Model}

We assume that the instantaneous hazard rate at time $t$ for individual $i_n^{(k)}$ is 
$$ \lambda_{i_n^{(k)}}(t|\bm{Z}_{i_n^{(k)}}, \varphi_k) = \lambda_0(t)  \exp{\left[\bm{\beta}^\top \bm{Z}_{i_n^{(k)}} + \varphi_k \right]},$$ where $\bm{Z}_{i_n^{(k)}}=(Z_{i_n^{(k)},1}, \dots, Z_{i_n^{(k)},p})^\top$ is a vector of $p$ covariates associated with the individual $i_n^{(k)}$, and $\varphi_k$ is the shared frailty associated with the spatial location $s_k$. The presence of a spatial cluster in the data results in an effect on the survival times of the spatial units involved. Hence, the effect of this cluster has been incorporated within the shared frailty: for each potential cluster $w$, $\varphi_k$ can be decomposed into a cluster effect $\alpha_w$ and $X_k$ being an effect specific to the spatial location $s_k$. Thus, the shared frailties $\varphi_k$ associated with the potential cluster $w$ can be rewritten as  $\varphi_k^{(w)} = \alpha_w \mathds{1}_{s_k \in w} + X_k$ where $\mathbb{E}[X_k]=0$. In this context the test hypotheses can be rewritten as $\mathcal{H}_0: \forall w \in \mathcal{W}, \alpha_w = 0$ (absence of cluster) and the alternative hypothesis associated with the potential cluster $w$ is $\mathcal{H}_1^{(w)}: \alpha_w \neq 0$ (presence of a cluster $w$ in which the individuals present atypical survival times).

Moreover, the spatial nature of the data requires taking into account a possible spatial correlation between the different spatial locations $s_k$, and thus between the $X_k$. This makes it possible to distinguish the effect of the cluster from the spatial correlation of unobserved factors at the scale of the spatial units.  Thus, we have considered the Conditional Autoregressive (CAR) model proposed by \cite{leroux} for the distribution of the $X_k$: $$X_k|X_{-k} \sim \mathcal{N}\left(\dfrac{\rho \dsum_{l=1}^{K} v_{k,l} X_l}{\rho \dsum_{l=1}^{K} v_{k,l} + 1 - \rho} ; \dfrac{\sigma_X^2}{\rho \dsum_{l=1}^{K} v_{k,l} + 1 - \rho} \right)$$ where $X_{-k}=\{X_1, \dots, X_{k-1}, X_{k+1}, \dots, X_K \}$, $v_{k,l}=1$ if $s_k$ and $s_l$ are adjacent (sharing a common boundary) and 0 otherwise, and $\rho \in [0;1]$ is the spatial correlation parameter. It should be noted that if there is no spatial correlation then the $X_k$'s are independent and identically distributed (i.i.d.) according to a normal distribution $\mathcal{N}(0,\sigma^2_X)$. Conversely, if $\rho =1$ (i.e., complete spatial correlation between the spatial units) then the $X_k$'s are distributed according to an Intrinsic CAR (ICAR) model \citep{besag1991bayesian}. 

The proposed method is decomposed into two steps. The first step (Section \ref{sec:estimationstep}) consists in estimating the shared frailties $\varphi_k$ as well as their spatial correlation parameter $\rho$. In a second step (Section \ref{sec:scanstep}), a scan procedure is developed and applied on the estimated shared frailties to identify spatial clusters of spatial units in which the $\varphi_k$ are significantly higher (higher risk) or lower (lower risk) than elsewhere. Lastly, the procedure for determining the statistical significance of the identified spatial clusters is described in Section \ref{sec:signifstep}.  

\subsubsection{Estimation of the $\varphi_k$ and $\rho$} \label{sec:estimationstep}

This first step consists in estimating the $\varphi_k$ and $\rho$ in a Bayesian framework using the integrated nested Laplace approximation (INLA) (see \cite{rue2009approximate} for details). 

The $\varphi_k$ are considered under both the $\mathcal{H}_0$ and $\mathcal{H}_1$ hypotheses. However, it should be noted that (i) both $X_k$ and $\rho$ do not depend on the clustering assumptions as they depend only on the spatial structure of the data and, (ii) only a single vector of $\varphi_k$ needs to be estimated in order to best fit the observed data. Therefore, the $\varphi_k$ must be estimated under the true hypothesis among $\mathcal{H}_0$ and the set of alternative hypotheses $\mathcal{H}_1^{(w)}$, i.e., the hypothesis under which the observations have been generated. Following the approach proposed by \cite{ahmed2021spatial}, we need to determine the ``best model'' among the candidate hypotheses ($\mathcal{H}_0$ and $\mathcal{H}_1^{(w)}$). For this purpose for each potential cluster $w \in \mathcal{W}$, we considered the Bayes Factor $\text{BF}^{(w)}$, defined as the marginal likelihood ratio between the model under $\mathcal{H}_1^{(w)}$ ($\mathcal{M}_1^{(w)}$) and the model under $\mathcal{H}_0$ ($\mathcal{M}_0$):
$$\text{BF}^{(w)} = \dfrac{\mathbb{P}\left[\left\{T_{i_n^{(k)}}, \delta_{i_n^{(k)}}, \bm{Z}_{i_n^{(k)}}, \mathds{1}_{i_n^{(k)} \in w} \right\}\big|\mathcal{M}_1^{(w)}\right]}{\mathbb{P}\left[\left\{T_{i_n^{(k)}}, \delta_{i_n^{(k)}}, \bm{Z}_{i_n^{(k)}}\right\}\big|\mathcal{M}_0\right]}.$$

Then, among all the models under $\mathcal{H}_1^{(w)}$ we select the ``best model'' $\mathcal{M}_1^{(w^*)}$ according to this criterion, i.e., the one associated with the potential cluster $w$ maximizing $\text{BF}^{(w)}$. Finally to decide if the estimates should be kept under $\mathcal{H}_0$ or under $\mathcal{H}_1^{(w^*)}$, we follow the rule of thumb proposed by \cite{jeffreys1998theory}: if $\text{BF}^{(w^*)}\ge 30$ then we keep the estimates (by the posterior mean) under $\mathcal{H}_1^{(w^*)}$, else we keep the estimates (by the posterior mean) under $\mathcal{H}_0$. This threshold of 30 corresponds to a ``very strong'' evidence for $\mathcal{H}_1^{(w^*)}$. 
Note that if the retained model is $\mathcal{M}_1^{(w^*)}$, the chosen estimate of $\varphi_k$ is $\varphi_k^*=\hat{\alpha}_{w^*}\mathds{1}_{s_k \in w^*} + \hat{X}_k$ and if the retained model is $\mathcal{M}_0$, $\varphi_k^* = \hat{X}_k$.

\subsubsection{Scan procedure} \label{sec:scanstep}

This section develops a scan procedure on the $\varphi_k^*$ to identify spatial clusters of spatial units in which the $\varphi_k^*$ are significantly higher (higher risk) or lower (lower risk) than elsewhere. Thus the hypotheses $\mathcal{H}_0$ and $\mathcal{H}_1^{(w)}$ are redefined in terms of the distribution of the $\varphi_k^*$ as follows:
 $$\mathcal{H}_0: \bm{\varphi}^{*} \sim \mathcal{N}(\alpha \mathds{1}, \sigma^{2(0)} A^{-1}) \text{ and }$$
 $$\mathcal{H}_1^{(w)}: \bm{\varphi}^{*} \sim \mathcal{N}(\alpha_w \mathds{1}_w + \alpha_{w^\mathsf{c}} \mathds{1}_{w^\mathsf{c}}, \sigma^{2(w)} A^{-1}), \ \alpha_w \neq \alpha_{w^\mathsf{c}}$$ 
 where $\bm{\varphi}^{*} = (\varphi_1^*, \dots, \varphi_K^*)^\top$, $\mathds{1}$ is the column vector composed only of 1,  $\mathds{1}_w$ and $\mathds{1}_{w^\mathsf{c}}$ are the column indicator vectors of $w$ and $w^\mathsf{c}$ respectively, and $A=\rho^*R + (1-\rho^*)\text{I}_K$ with $R$ the square matrix composed of the elements
 $$R_{k,l} = \left\{ \begin{array}{ll}
\dsum_{j=1}^K v_{k,j} & \text{ if } k = l \\
- v_{k,l} & \text{ otherwise}
\end{array}  \right..$$
Note that these assumptions are equivalent to considering both under $\mathcal{H}_0$ and $\mathcal{H}_1^{(w)}$ the same variance-covariance structure as with the CAR model considered previously (proof in Supplementary Material \ref{sec:leroux}), and since $w \cap w^\mathsf{c}=\emptyset$, to assume different frailty means in $w$ and $w^\mathsf{c}$ under $\mathcal{H}_1^{(w)}$ (respectively $\alpha_w$ and $\alpha_{w^{\mathsf{c}}}$).

The unknown parameters $\alpha$, $\sigma^{2(0)}$, $\alpha_w$, $\alpha_{w^\mathsf{c}}$ and $\sigma^{2(w)}$ are estimated by their maximum likelihood estimators (proofs are available in the Appendix \ref{appendix:proofs}): 
\begin{align*}
\hat{\alpha} &= \dfrac{\mathds{1}^\top A \bm{\varphi}^{*}}{\mathds{1}^\top A \mathds{1}}, \\
\widehat{\sigma^{2(0)}} &= \dfrac{1}{K} [\bm{\varphi}^{*\top} A \bm{\varphi}^{*} - 2 \hat{\alpha} \mathds{1}^\top A \bm{\varphi}^{*} + \hat{\alpha}^2 \mathds{1}^\top A \mathds{1}], \\
\hat{\alpha}_{w^\mathsf{c}} &= \left[ \mathds{1}_{w^\mathsf{c}}^\top A \mathds{1}_{w^\mathsf{c}} - \dfrac{\mathds{1}_w^\top A \mathds{1}_{w^\mathsf{c}} \mathds{1}_w^\top A \mathds{1}_{w^\mathsf{c}} }{\mathds{1}_w^\top A \mathds{1}_w}\right]^{-1} \left[\mathds{1}_{w^\mathsf{c}}^\top A \bm{\varphi}^{*} - \dfrac{\mathds{1}_w^\top A \bm{\varphi}^{*} \mathds{1}_w^\top A \mathds{1}_{w^\mathsf{c}} }{\mathds{1}_w^\top A \mathds{1}_w} \right], \\ 
\hat{\alpha}_w &= \dfrac{\mathds{1}_w^\top A \bm{\varphi}^{*} - \hat{\alpha}_{w^\mathsf{c}} \mathds{1}_w^\top A \mathds{1}_{w^\mathsf{c}}}{\mathds{1}_w^\top A \mathds{1}_w} \text{ and }\\ \widehat{\sigma^{2(w)}} &= \dfrac{1}{K} [\bm{\varphi}^{*} - \hat{\alpha}_w \mathds{1}_w - \hat{\alpha}_{w^\mathsf{c}} \mathds{1}_{w^\mathsf{c}} ]^\top A [\bm{\varphi}^{*} - \hat{\alpha}_w \mathds{1}_w - \hat{\alpha}_{w^\mathsf{c}} \mathds{1}_{w^\mathsf{c}}].
\end{align*}

Then the log-likelihood function under $\mathcal{H}_0$ has the following expression:
\begin{align*}
\ell_{\mathcal{H}_0}(\hat{\alpha}, \widehat{\sigma^{2(0)}}) &= - \dfrac{K}{2} \ln{[2\pi]} - \dfrac{1}{2} \ln{|A^{-1}|} - \dfrac{K}{2} \ln{[\widehat{\sigma^{2(0)}}]} - \dfrac{K}{2},
\end{align*}
while the log-likelihood function associated with $\mathcal{H}_1^{(w)}$ can be expressed as follows:
\begin{align*}
\ell_{\mathcal{H}_1}(\hat{\alpha}_w, \hat{\alpha}_{w^\mathsf{c}}, \widehat{\sigma^{2(w)}}) &= - \dfrac{K}{2} \ln{[2\pi]} - \dfrac{1}{2} \ln{|A^{-1}|} - \dfrac{K}{2} \ln{[\widehat{\sigma^{2(w)}}]} - \dfrac{K}{2}.   
\end{align*}

Thus the log-likelihood ratio associated with the potential cluster $w$ is 
\begin{align*}
LLR^{(w)} &= \ell_{\mathcal{H}_1}(\hat{\alpha}_w, \hat{\alpha}_{w^\mathsf{c}}, \widehat{\sigma^{2(w)}}) -  \ell_{\mathcal{H}_0}(\hat{\alpha}, \widehat{\sigma^{2(0)}}) \\
&= \dfrac{K}{2}\left[\ln{\dfrac{\widehat{\sigma^{2(0)}}}{\widehat{\sigma^{2(w)}}}}\right]
\end{align*}

and finally the spatial scan statistic can be defined as
$$ \Lambda = \underset{w \in \mathcal{W}}{\max} \ LLR^{(w)}.$$
The most likely cluster (MLC) is then defined as 
$$ \text{MLC} = \underset{w \in \mathcal{W}}{\arg \max} \ LLR^{(w)}.$$

\subsubsection{Statistical significance} \label{sec:signifstep}

Once the MLC has been detected, its statistical significance must be evaluated. However, the distribution of $\Lambda$ does not have a closed form under $\mathcal{H}_0$, so in the literature this distribution is usually approximated by a Monte-Carlo procedure \citep{dwass}. Two main methods can be distinguished according to the presence (or not) of a distributional hypothesis made on the data.  The first method consists in generating data sets under $\mathcal{H}_0$, thus requiring a distributional assumption on the data \citep{kulldorff1997spatial}. The second method, namely random labelling, consists in randomly permuting the observations among the spatial locations \citep{normalkulldorff}. In the present case, this method is not applicable because the permutations of the observations do not preserve the spatial correlation of the observations. Therefore, the approximation of the distribution of $\Lambda$ under $\mathcal{H}_0$ was performed using the first method: since we have assumed a distribution on the $\varphi_k$, one generates $M$ data sets under $\mathcal{H}_0$ \textit{via} $\hat{\alpha}$ and $\widehat{\sigma^{2(0)}}$ which correspond respectively to the estimators of the mean and variance of the $\varphi_k$ under $\mathcal{H}_0$. For each generated data set $m$, $1\le m\le M$, one compute the associated spatial scan statistic $\Lambda^{(m)}$, leading to an approximation of the distribution of $\Lambda$ under $\mathcal{H}_0$. Finally, the p-value associated with the MLC is estimated by $$\hat{p} = \dfrac{1 + \dsum_{m=1}^M \mathds{1}_{\Lambda^{(m)} \ge \Lambda}}{M+1} .$$

\section{Simulation studies} \label{sec:simulation}

\cite{huang2007spatial} and \cite{cook2007spatial} proposed spatial scan statistics for time-to-event data indexed in space. However they supposed that the individuals are independent which is a strong and not really realistic hypothesis since individuals located in the same spatial unit can be correlated. Thus, in a first simulation study (Section \ref{sec:simuclassical}), we investigated  the impact of the presence of intra-spatial unit correlation on the type I error of the methods proposed by \cite{huang2007spatial} and \cite{cook2007spatial}. \\
In Section \ref{sec:simucar}, we conducted two simulation studies. The first one (Section \ref{subsec:quality_estimates}) aimed at evaluating the ability of our method to correctly estimate both the spatial correlation parameter and the cluster effect. The second study (Section \ref{subsec:spatial_dep}) aimed, in the presence of spatial correlation, to evaluate the performance of our approach in the context of cluster detection, but also to compare the performance of our method to those of its two particular versions: i.i.d. ($\rho=0$) and ICAR ($\rho=1$).\\
Lastly, Section \ref{sec:simucensure} presents a simulation study that investigated the performance of our approach in the presence of different levels of censoring of time-to-event data.

\subsection{Impact of an intra-spatial unit correlation on the type I error of standard methods} \label{sec:simuclassical}

In this simulation study, we evaluated the type I errors of classical spatial scan statistics for cluster detection in survival data, namely the exponential model proposed by \cite{huang2007spatial} and the method based on a log-rank test proposed by \cite{cook2007spatial}, in the presence of an intra-spatial unit correlation.

\subsubsection{Design of the simulation}

\noindent We considered 1690 individuals distributed in 169 spatial units corresponding to administrative subdivisions of northern France, located by their centroid. We defined a spatial cluster $w$ (characterized by $\alpha$) composed of 135 individuals located in 14 contiguous spatial units (the green area on Figure \ref{fig:map_cluster} in Supplementary Materials \ref{appendix:design}). \\

\noindent We considered the following simulation model: for the individual $i_n^{(k)}$ in the spatial unit $s_k$,  $$\lambda_{i_n^{(k)}}(t|\varphi_k) = \lambda_0(t) \exp{[\varphi_{k}]},$$
with $\lambda_0(t) = \frac{1}{2}$ which results in an exponential model.
The event times were simulated by the inverse transform sampling: for each individual $i_n^{(k)}$, we generated a uniformly distributed random number $u_{i_n^{(k)}}$ on $[0;1]$, which allowed to generate a survival time $T_{i_n^{(k)}}$ by $T_{i_n^{(k)}} = \underset{t > 0}{\inf} \ 1-S_{i_n^{(k)}}(t) > u_{i_n^{(k)}}$. Note that this results in $T_{i_n^{(k)}} = -2\ln{[1-u_{i_n^{(k)}}]} \exp{[-\varphi_k]}$.

\noindent The $\varphi_{k}$ were defined by the vector $\bm{\varphi} = (\varphi_1, \dots, \varphi_K)^\top$ such that
$$\bm{\varphi} \sim \mathcal{N}(\alpha \mathds{1}_{w}, \sigma^2 [\rho R + (1-\rho) I_K]^{-1}),$$
where $\mathds{1}_{w}$ is the column indicator vector of $w$. \\

Here we focused the analysis on the type I errors ($\alpha=0$) of the exponential model \citep{huang2007spatial} and the log-rank test method \citep{cook2007spatial} in the presence of a non spatially correlated ($\rho=0$) shared frailty for the frailty variance $\sigma^2$ varying from 0.001 to 0.101 by increments of 0.010. \\
For each value of $\sigma^2$, 100 data sets were simulated. The statistical significance of the MLC was evaluated as proposed by the authors of the two methods, i.e., by using 999 permutations of the data. The type I error was set to 0.05.

\subsubsection{Results}
\begin{figure}[h!]
\centering
\includegraphics[width = \linewidth]{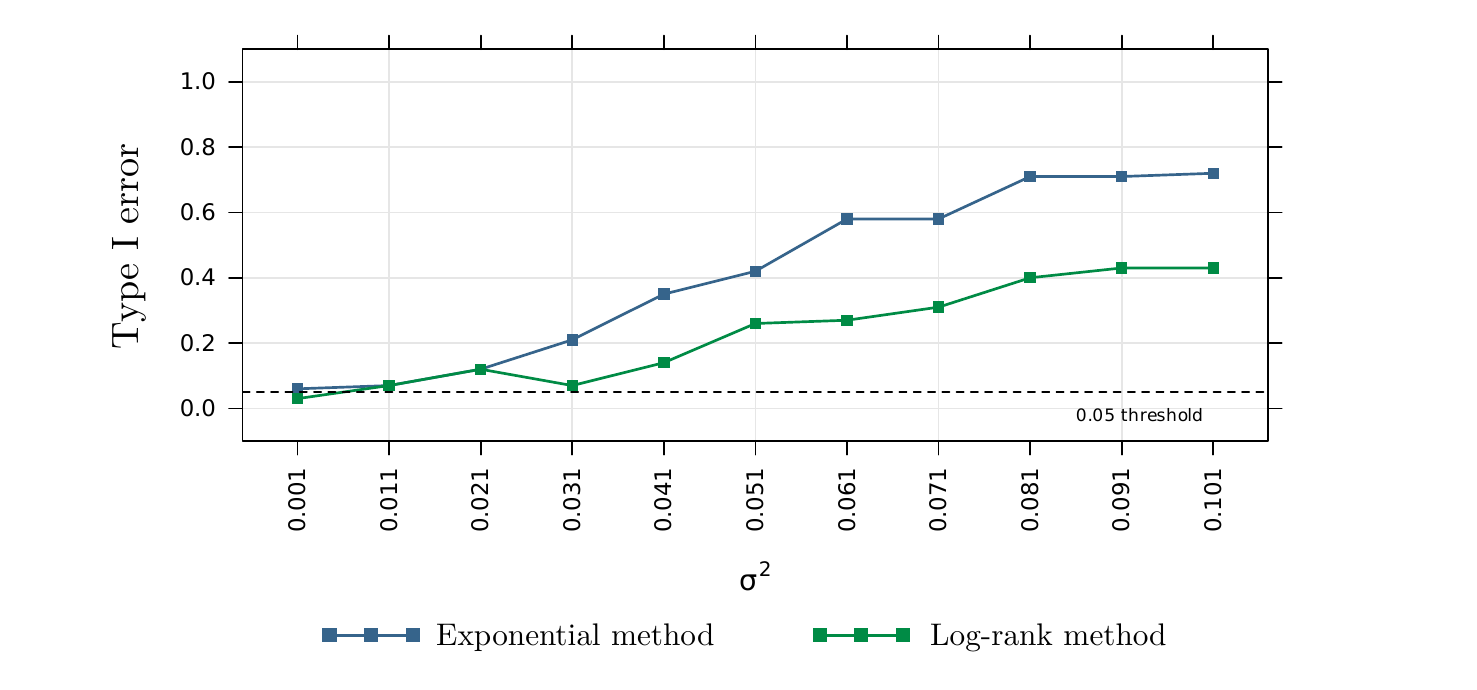}
\caption{Type I error of the exponential method of \cite{huang2007spatial} and the log-rank one \citep{cook2007spatial} according to the presence of different levels of intra-unit spatial correlation (characterized by the values of the simulated shared frailty variance $\sigma^2$). 
}
\label{fig:k_lg}
\end{figure}

Figure \ref{fig:k_lg} shows the values of the type I error as a function of the values of $\sigma^2$. We can notice that the type I error increases when the values of $\sigma^2$ increase, showing that the nominal level is not preserved.  This can be explained by the fact that, under the hypothesis $\mathcal{H}_0$ of absence of cluster, the increase of $\sigma^2$ leads directly to the increase of the variance of $X_k$. Since the two standard models do not incorporate a shared frailty, the identification of false-positive spatial clusters is essentially due to the intra-unit spatial correlation (i.e, the variance of $X_k$). 

\subsection{Performance evaluation of the proposed method} \label{sec:simucar}

Here two simulation studies were conducted to evaluate the performance of our approach. The first one (Section \ref{subsec:quality_estimates}) aimed at assessing the ability of our method to accurately estimate both the spatial correlation parameter and the cluster effect. The second study (Section \ref{subsec:spatial_dep}) aimed at evaluating the performance of our method in the context of cluster detection, and comparing it to two particular versions of the model: the one assuming no spatial correlation (i.i.d. frailty model) and the one assuming a complete spatial correlation (ICAR frailty model), in the presence of spatial correlation.

\subsubsection{Design of the simulation studies}
The design of these simulation studies is very similar to that presented in Section \ref{sec:simuclassical}. The only differences are that since we wanted to investigate the impact of spatial correlation on cluster detection, we set $\sigma^2$ to 1 and we considered several values for the parameters controlling the spatial correlation $\rho \in \{0;0.2;0.4;0.6;0.8\}$ and the cluster effect $\alpha \in \{0;0.5;1;1.5;2\}$. Note that $\alpha=0$ was considered to evaluate the maintenance of the type I error.

For each value of the spatial correlation parameter $\rho$ and each value of $\alpha$, 100 data sets were simulated.  
The statistical significance of the MLC was evaluated through 999 generations of the data under $\mathcal{H}_0$ (see Section \ref{sec:signifstep} for more details) and the type I error was set to 0.05. \\
\noindent The performances were measured through 4 criteria: the power, the true positive rate, the false positive rate and the positive predictive value. The power was estimated as the proportion of simulations leading to the rejection of $\mathcal{H}_0$, depending on the type I error. Among the simulated data sets leading to the rejection of $\mathcal{H}_0$, the true positive rate was defined as the average proportion of individuals correctly detected among the individuals in $w$, the false positive rate as the average proportion of individuals in $w^\mathsf{c}$ that were included in the detected cluster, and the positive predictive value corresponded to the proportion of individuals in $w$ within the detected cluster. 

Since the estimations of the $\varphi_k$ and $\rho$ were performed in a Bayesian framework, we considered the following Leroux CAR prior for $X_k$: $\bm{X} \sim \mathcal{N}(\bm{0}, \sigma^2[\rho R + (1-\rho)I_K]^{-1})$, with 
 a $\bm{\beta}(1;1)$ prior for the spatial correlation parameter $\rho$ and a $\Gamma(10^{-3},10^{-3})$ prior for the precision $1/\sigma^2$. For $\alpha_w$, we chose a non informative prior $\mathcal{N}(0,10^3)$. \\
Lastly, for the baseline hazard $\lambda_0$, the observation times were divided into $n_T$ time intervals. Here $n_T$ was set to be the number of unique times divided by 20. Then, $\lambda_0$ was supposed to be constant in each time interval, and for each interval $I$ we supposed $\lambda_0 = \exp(c_I)$. We chose a Gaussian prior on the $c_I$ increments with precision $\tau$ such that $\tau \sim \Gamma(10^{-3}, 10^{-3})$: $\Delta c_I = c_I - c_{I-1} \sim \mathcal{N}(0, \tau^{-1})$.

\subsubsection{Evaluation of the estimates of $\rho$ and $\alpha_w$} \label{subsec:quality_estimates}
Section \ref{sec:estimationstep} presented
the estimation step of the $\varphi_k$. Briefly, it consists in choosing either the estimates under the best hypothesis $\mathcal{H}_1^{(w)}$: $\mathcal{H}_1^{(w^*)}$ (in this case the estimates are $\varphi_k^*=\hat{\alpha}_{w^*} \mathds{1}_{s_k \in w^*} + \hat{X}_k$) or the estimates under $\mathcal{H}_0$ ($\varphi_k^*=\hat{X}_k$). 
This section focuses on the bias of the estimates obtained for the spatial correlation parameter ($\rho^*$), as well as for the cluster effect ($\hat{\alpha}_{w^*}$). Note that for the latter we only considered the simulations that did not retain $\mathcal{H}_0$ (since otherwise no estimate $\hat{\alpha}_{w^*}$ was available). Thus the obtained estimates were compared to the true values of the spatial correlation parameter and the cluster effect.

\begin{figure}[h!]
\begin{minipage}{\linewidth}
(a)\\
\includegraphics[width=\textwidth]{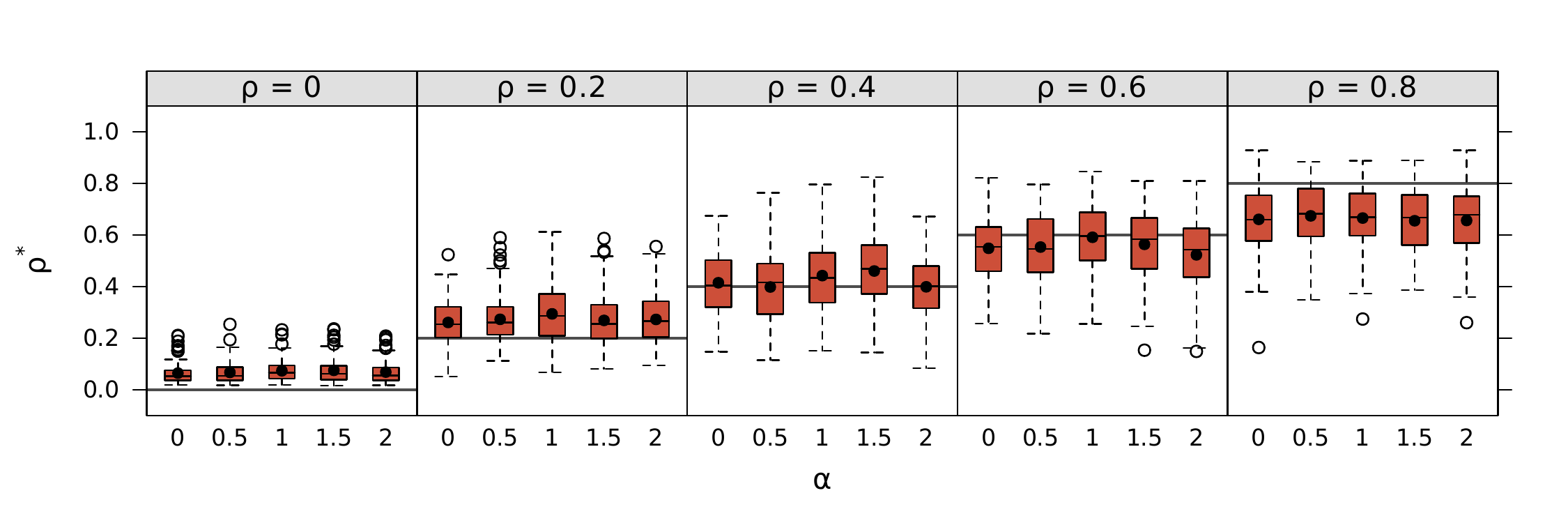}
\end{minipage}
\begin{minipage}{\linewidth}
(b) \\
\includegraphics[width=\textwidth]{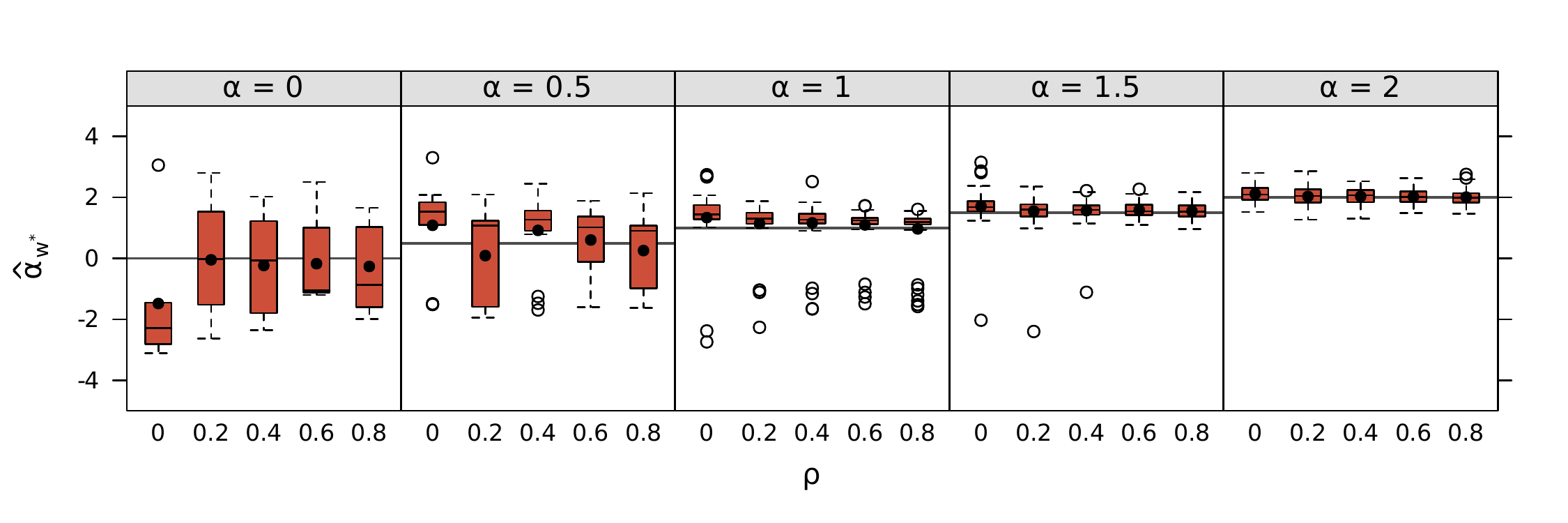}
\end{minipage}
\caption{Simulation study: the selected $\rho^*$ according to the parameters $\rho$ and $\alpha$ (panel (a)) and $\hat{\alpha}_{w^*}$ obtained with INLA when we select $\mathcal{H}_1$ according to the Bayes Factor criterion (panel (b)). The main horizontal lines correspond to the true value of the parameters $\rho$ and $\alpha$ in panels (a) and (b) respectively and the black points represent the mean estimates obtained.}
\label{fig:boxplot_alphasrhos_car}
\end{figure}

Figure \ref{fig:boxplot_alphasrhos_car} shows the selected $\rho^*$ according to the parameters $\rho$ and $\alpha$ as well as the estimations $\hat{\alpha}_{w^*}$ with INLA when we select $\mathcal{H}_1$ according to the Bayes Factor criterion.

The cluster effect is well estimated by our approach when the simulated values of $\alpha$ are equal to 1, 1.5 and 2. However, it must be pointed that for values of $\alpha$ equal to 0 and 0.5, the cluster effect appears to be poorly estimated. This point arises from the fact that, for these values, our approach rarely selected $\mathcal{H}_1$ and, therefore, few estimates were made.

The parameter $\rho$ is globally well estimated although it is slightly overestimated for $\rho=0$ and slightly underestimated for $\rho=0.8$.

\subsubsection{Impact of spatial correlation on cluster detection} \label{subsec:spatial_dep}
Here the performance of the proposed method was evaluated in the context of cluster detection. 
Two particular versions of the method were also considered to investigate the impact of not accounting for spatial correlation although taking into account a potential intra-spatial unit correlation (i.i.d. model, $\rho=0$), or of accounting for it without adjusting its intensity by considering it to be complete (ICAR model, $\rho=1$).

Note that for the ICAR model, $\rho=1$ leads to the non-invertibility of the matrix $A$ and therefore it was not possible to
generate data under $\mathcal{H}_0$ to estimate the p-value associated with the most likely cluster (see Section \ref{sec:signifstep} for more details). To overcome this problem,the value of the spatial correlation was fixed at 0.999 instead of 1 in the scan procedure (Section \ref{sec:scanstep}) for the ICAR model.

\begin{figure}[h!]
    \centering
    \includegraphics[width=0.9\textwidth]{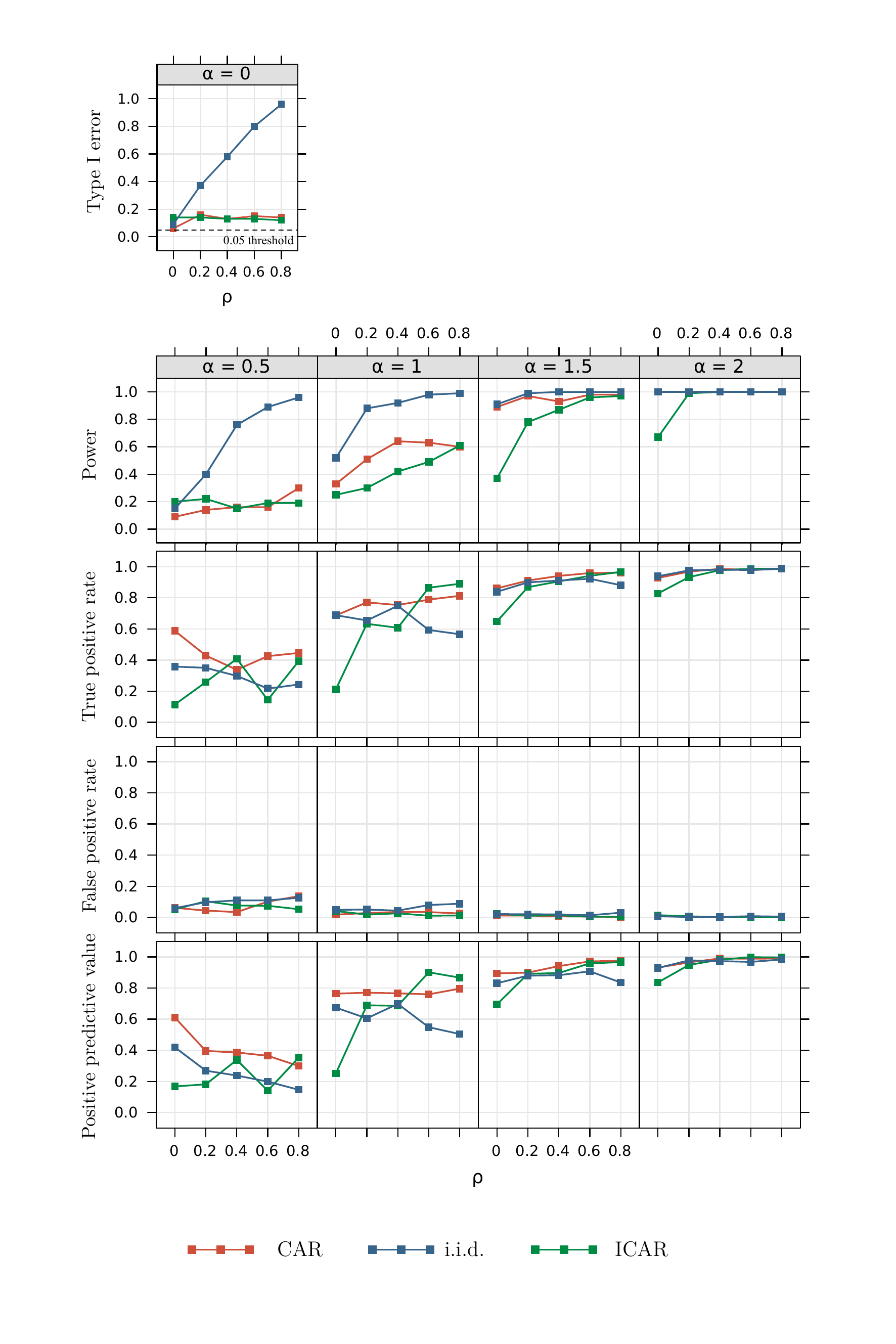}
    \caption{Simulation study: Comparison of the type I error, power curves, true positive and false positive rates and positive predictive values for the CAR, ICAR and i.i.d. models. $\alpha$ is the parameter that controls the cluster intensity and $\rho$ controls the spatial correlation.}
    \label{fig:perf_iid_icar_full}
\end{figure}

Figure \ref{fig:perf_iid_icar_full} shows the type I error, the power curves, true positive rates, false positive rates and positive predictive values obtained with our method, as well as with its two special cases: i.i.d. and ICAR. 

For the Leroux CAR model, the performances are relatively stable according to the values of $\rho$ although the type I error is slightly higher than the fixed threshold of 5\% (except for $\rho=0$ but this is due to the fact that when $\rho=0$, $\rho^*$ slightly overestimates $\rho$ (Figure \ref{fig:boxplot_alphasrhos_car}) which makes the method quite conservative in this case). Remark that the true and false positive rates and the positive predictive values appear to be less stable according the values of $\rho$ for $\alpha=0.5$. This is due to the fact that these indicators are only computed on the simulations rejecting $\mathcal{H}_0$, which are few in this case. 

The i.i.d. model fails to maintain a reasonable type I error when $\rho$ increases. Moreover, power according to the values of $\rho$ was less stable than in the CAR model.

The ICAR model tended to absorb the cluster effect in the spatial correlation parameter $\rho$. This is particularly the case when the true value of $\rho$ is low. Thus the type I errors remain reasonable but the power tended to decrease when $\rho$ decreases. 

The false positive rates were very low for the three approaches. However, the true positive rates and the positive predictive values were lower for the i.i.d. and the ICAR models than the CAR model.\\

Note that the performance of our method was also investigated when considering other thresholds for the Bayes Factor (i.e., 3, 10 and 100 which correspond respectively to ``substantial'', ``strong'' and ``decisive'' evidences for $\mathcal{H}_1^{(w^*)}$ \citep{jeffreys1998theory}). The results are presented in Figure \ref{fig:perf_car_thresholds} in the Appendix \ref{appendix:otherbf}.

\subsection{Influence of censoring} \label{sec:simucensure}

In this section, a simulation study was conducted to evaluate the performance of our approach in the presence of different levels of censoring in the data.

\subsubsection{Design of the simulation}
For computational time constraints, the design of the simulation is slightly different from the previous studies: we considered 940 individuals distributed in the 94 French \textit{départements} located by their centroid. The simulated cluster contains 73 individuals in the 8 \textit{départements} of \textit{Île de France} region (see the green area in Figure \ref{fig:map_cluster94} in Supplementary Materials \ref{appendix:design}).

The data were generated in the same way as in Section \ref{sec:simucar} except that a given proportion of observations were censored. In order to simulate different censoring percentages (10\%, 20\%, 30\% and 40\%), administrative censoring was considered following \cite{montez2017guidelines}. Briefly, the end of the study was determined so that the right proportion of the observations was censored.

For each value of the spatial correlation parameter $\rho$, each value of $\alpha$, and each censoring percentage, 100 data sets were simulated.  
The statistical significance of the MLC was evaluated through 999 generations of the data under $\mathcal{H}_0$ (see Section \ref{sec:signifstep} for more details) and the type I error was set to 0.05. \\
\noindent The performances were measured through the same 4 criteria than in Section \ref{sec:simucar}: the power, the true positive rate, the false positive rate and the positive predictive value. \\ 
Note that in this simulation study the same \textit{a priori} distributions as in Section \ref{sec:simucar} were considered.

\subsubsection{Results}
\begin{figure}[h!]
    \centering
    \includegraphics[width=0.9\textwidth]{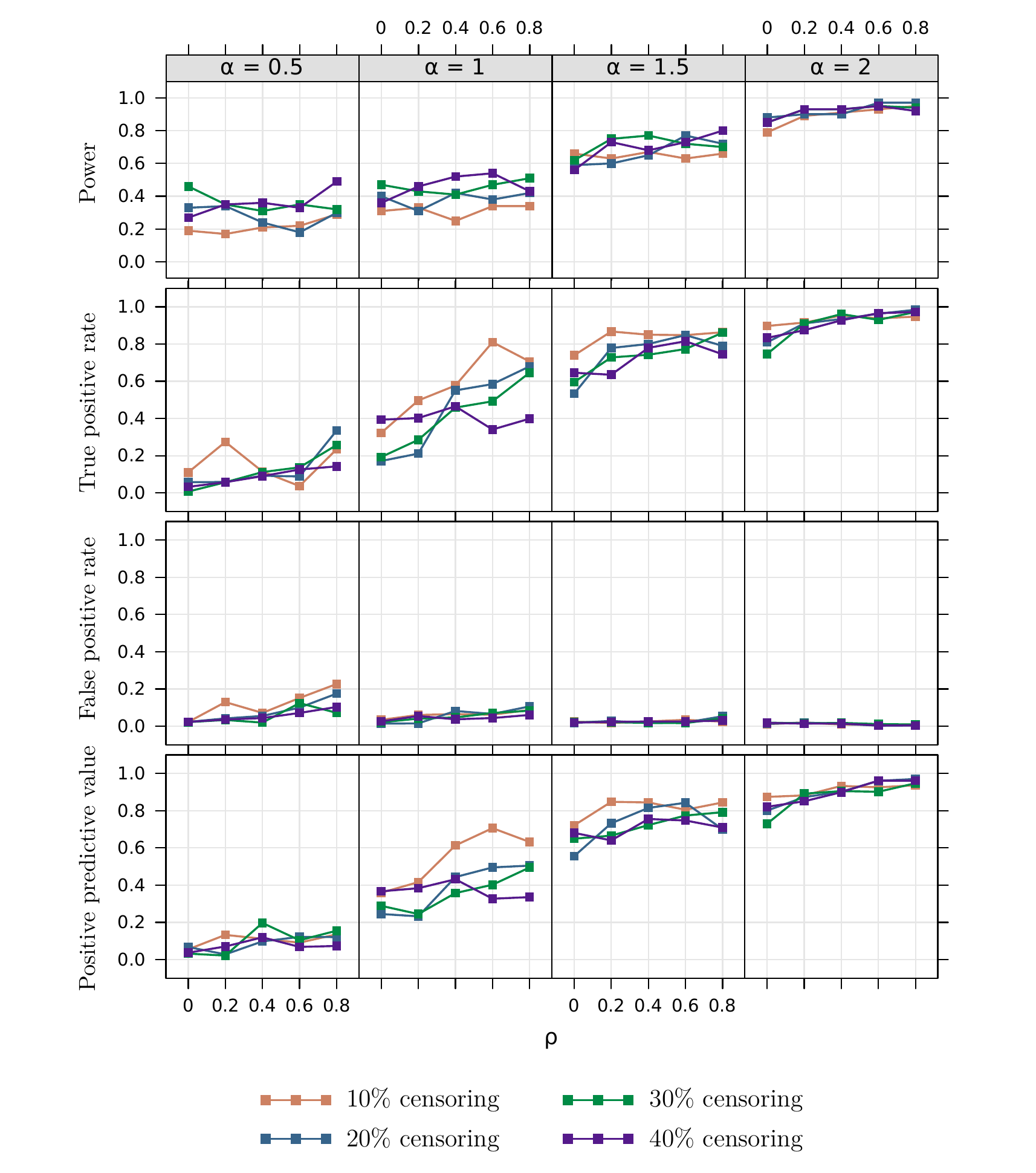}
    \caption{Simulation study: comparison of the power curves, true positive and false positive rates and positive predictive values according to different percentages of censored observations. $\alpha$ is the parameter that controls the cluster intensity and $\rho$ controls the spatial correlation.}
    \label{fig:perf_censure_full}
\end{figure}

The results of this simulation study are shown in Figure \ref{fig:perf_censure_full}.  First, we can observe that the power of our method increases when the proportion of censoring increases. This point is also observed for the type I error (see Figure \ref{fig:typeIcensure} in Supplementary Materials \ref{appendix:typeIcensure}). Although the latter remains stable and close to the nominal type I error (whatever the values of $\rho$) when 10\% of the observations are censored, it tends to move away from the nominal type I error as the censoring percentage increases.\\
False positive rates remain very low regardless of the censoring rate but true positive rates and positive predictive values decrease as the censoring rate increases. However, the impact of censoring on the performance indicators decreases when the intensity $\alpha$ of the cluster increases.

\section{Application to epidemiological data} \label{sec:realdata}

\subsection{ESRD mortality and related confounding factors}
\label{application:ESRD_descr}
We considered data provided by the French renal epidemiology and information network (REIN) registry on end-stage renal disease (ESRD) in northern France between 2004 and 2020. The methodology of the REIN registry has been described elsewhere \citep{Couchoud_2005}. Here, we focused on the analysis of the following endpoint: the mortality measured by the survival time after initiation of dialysis in the population of ESRD patients aged 70 years and older. This population is characterized by (i) a high mortality rate, thus ensuring a high number of observed deaths and, (ii) a low rate of kidney transplantation, thus minimizing the effect of this known competing risk of death among ESRD patients \citep{Hallan_2012,ayav2016competing}. The data consist of 6071 individuals for whom the exact time to survival after initiation of dialysis is not known in 17\% of cases. These censored observations are either patients still alive at the end of the study (15.7\%), or patients lost to follow-up (0.7\%), or patients having undergone a renal transplantation (in this case the censoring time corresponds to the date of transplantation) (0.6\%). The geographical region studied (northern France, \textit{Nord-Pas-de-Calais} region) is divided into 80 \textit{cantons} (a French administrative subdivision) and each individual was associated with a \textit{canton}, according to his place of residence.

We also considered 18 variables measured at the individual level that are known to be confounders of ESRD patient survival \citep{couchoud2015development, fu2021timing}. Thus, spatial cluster detection was adjusted by introducing the following confounders into each model as covariates: age (in years), sex, body mass index (in $\text{kg/m}^{2}$), type of nephropathy (polycystic, primitive glomerulonephritis, hypertension or vascular, diabetic, pyelonephritis, other), number of cardiovascular comorbidities (none, one, at least two), 
mobility (autonomous walking, need for a third party, total disability), hemoglobin level (in g/dL), albumin level (in g/dL), dialysis method (haemodialysis or peritoneal dialysis), glomerular filtration rate (below 7, between 7 and 10 or over 10 mL/min/1.73m$^2$), year of treatment initiation (2004-2009, 2010-2015 or 2016-2020), whether or not the treatment was initiated urgently, and the absence or presence of diabete, chronic respiratory disease, respiratory assistance, cirrhosis, severe behavioral disorder and active malignancy. Details of these confounding factors are available in Supplementary Material \ref{appendix:application}.\\

\subsection{Spatial clusters detection}
In order to detect spatial clusters of atypical survival times in ESRD patients, 5 models were considered: the exponential model (Model 1) proposed by \cite{huang2007spatial}, the log-rank method proposed by \cite{cook2007spatial} (Model 2) and the method presented in this paper based on a Cox model considering three types of shared frailty: i.i.d. ($\rho=0$) (Model 3), CAR ($\rho\in ]0,1[$) (Model 4) and ICAR ($\rho=1$) (Model 5).\\

Each model was used to detect spatial clusters of shorter (or longer) survival times compared to the rest of the studied region, among the ESRD patients.
To adjust for the confounders in Model 1 we used an exponential regression method to obtain adjusted survival times as proposed by \cite{huang2007spatial}. Regarding Model 2, we followed the approach proposed by \cite{jung2009generalized, ahmed2020functional} that consists in estimating the coefficients associated with the confounders in the model under $\mathcal{H}_0$ and then fixing their value in the scan statistic proposed by \cite{cook2007spatial}. Regarding Models 3 to 5 (the shared frailty models), we also followed this approach by fixing under each alternative hypothesis $\mathcal{H}_1^{(w)}$ the coefficients associated with the confounding factors to their estimated values in the model under $\mathcal{H}_0$ in the estimation step of the $\varphi_k$ (Section \ref{sec:estimationstep}).

In order to provide an indicator of the risk associated with the clusters, which is independent of the model considered, we estimated the Hazard Ratio (HR) associated with each cluster in a classical Cox model adjusted for the confounding factors.

The MLC was considered, together with secondary clusters that had a high value of $\Lambda$ and did not cover the MLC \citep{kulldorff1997spatial}. The statistical significance of the detected spatial clusters was evaluated by performing 999 Monte-Carlo simulations, with a type I error of 0.05. \\

\subsection{Results}

\begin{figure}[h!]
\centering
\includegraphics[width=\textwidth]{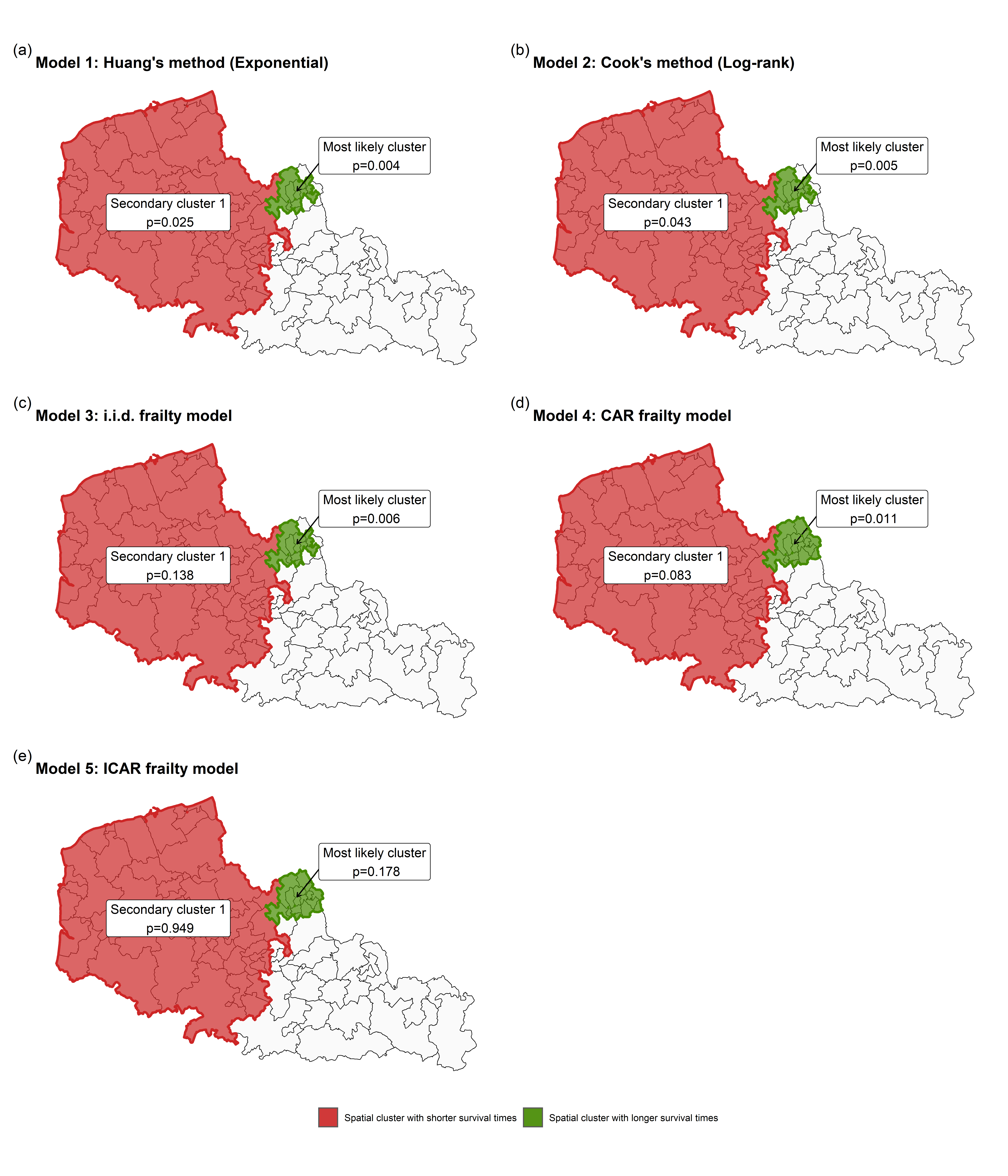}
\caption{Spatial clusters detected by the method proposed by \cite{huang2007spatial} (Model 1 (exponential), panel(a)), the method of \cite{cook2007spatial} (Model 2 (Log-rank), panel (b)) and those highlighted by the shared frailty models (Model 3 (i.i.d.), panel (c); Model 4 (CAR), panel (d); Model 5 (ICAR), panel (e)) after adjusting on confounding factors. Spatial clusters in green indicate longer survival times for ESRD patients compared to the rest of the studied region. Those in red indicate shorter survival times for ESRD patients.}
\label{fig:map_res}
\end{figure}

The spatial clusters detected by each of the 5 models (Exponential, Log-rank, i.i.d., CAR and ICAR frailty) are presented in Figure \ref{fig:map_res}. Detailed information on the spatial clusters is presented in Table \ref{tab:descriptionclusters}.\\ 

Both the exponential model (Model 1, panel (a) in Figure \ref{fig:map_res}) and the method based on the log-rank test (Model 2, panel (b) in Figure \ref{fig:map_res}) identified the same two statistically significant spatial clusters. The MLC, located in the northeast of the region (green color) presented a close level of significance ($\hat{p}=0.004$ and $\hat{p}=0.005$ respectively) and showed longer survival times than in the rest of the geographical area studied (HR = 0.84 for both models). The first secondary cluster, located in the western part of the region (red color), also had a close level of significance ($\hat{p}=0.025$ and $\hat{p}=0.043$ respectively) and was characterized by shorter survival times (HR=1.13 for both methods).

The i.i.d. frailty model (Model 3, panel (c)) identified the same statistically significant MLC as the exponential model and the method based on the log-rank test ($\hat{p}=0.006$). The CAR (Model 4, panel (d)) and ICAR (Model 5, panel (e)) models both detect the same MLC which contains three more spatial units than the MLC of the other models. It is also characterized by longer survival times (HR=0.86).
Although it is statistically significant for the CAR model, it is not statistically significant for the ICAR model ($\hat{p}=0.011$ and $\hat{p}=0.178$ respectively).
The first secondary cluster of the three frailty models is the same as the first secondary cluster of the exponential model and the method based on the log-rank test. However it is not statistically significant with the frailty models ($\hat{p}$ = 0.138 for the i.i.d. frailty model, $\hat{p}$ = 0.083 for the CAR frailty model and $\hat{p}$ = 0.949 for the ICAR frailty model respectively).

The small differences between the classical spatial scan statistics methods \citep{huang2007spatial,cook2007spatial} and the three shared frailty models developed here can be explained by the low variance of the shared frailties (see Figure \ref{fig:posteriors} in Supplementary Materials \ref{appendix:application} for the posterior distribution of $\sigma^2$ with each model).

\begin{table}[h!]
\centering
\caption{Description of the statistically significant spatial clusters detected by the method proposed by \cite{huang2007spatial} (Model 1 (exponential)), the method of \cite{cook2007spatial} (Model 2 (Log-rank)) and those detected by the shared frailty models (Model 3 (i.i.d.), Model 4 (CAR) and Model 5 (ICAR)) after adjusting for confounding factors.}

\begin{small}
\begin{tabular}{ccccccc}
\toprule
\multirow{2}{*}{\textbf{Model}} & \multirow{2}{*}{\textbf{Cluster}} & \multirow{2}{*}{\textbf{p-value}} & \textbf{Number of} & \textbf{Number of} & \textbf{Number of} & \textbf{Hazard} \\
 & & & \textbf{spatial units} & \textbf{individuals} & \textbf{events} & \textbf{Ratio}$^1$ \\
\midrule
\multirow{1}{*}{Model 1} & MLC & 0.004 & 10 & 1091 & 890 & 0.84 \\
(Exponential) & Secondary cluster 1 & 0.025 & 43 & 2632 & 2163 & 1.13 \\
\midrule
\multirow{1}{*}{Model 2} & MLC & 0.005 & 10 & 1091 & 890 & 0.84 \\
(Log-rank) & Secondary cluster 1 & 0.043 & 43 & 2632 & 2163 & 1.13 \\
\midrule
\multirow{1}{*}{Model 3} & MLC & 0.006 & 10 & 1091 & 890 & 0.84 \\
(i.i.d. frailty) & Secondary cluster 1 & 0.138 & 43 & 2632 & 2163 & 1.13 \\
\midrule
\multirow{1}{*}{Model 4} & MLC & 0.011 & 13 & 1346 & 1094 & 0.86 \\
(CAR frailty) & Secondary cluster 1 & 0.083 & 43 & 2632 & 2163 & 1.13 \\
\midrule
\multirow{1}{*}{Model 5} & MLC & 0.178 & 13 & 1346 & 1094 & 0.86 \\
ICAR frailty & Secondary cluster 1 & 0.949 & 43 & 2632 & 2163 & 1.13 \\
\bottomrule
\end{tabular}

$^1$ The Hazard Ratio (HR) is computed with a Cox model with adjustment for confounders.
\end{small}
\label{tab:descriptionclusters}
\end{table}

\section{Discussion} \label{sec:discussion}
Here we developed a new spatial scan statistic for survival data indexed in space that allows to take into account both a potential intra-spatial unit correlation and spatial correlation, as well as to adjust the cluster detection on confounding factors. This method is based on a Cox model that includes a spatially structured shared frailty distributed according to a Leroux CAR model. \\

In a simulation study we showed that, in the presence of intra-spatial unit correlation, the existing methods \citep{cook2007spatial, huang2007spatial} face a huge increase of the type I error. Thereafter, the performance of the CAR model was evaluated in the context of cluster detection and compared to two particular versions of the latter: the i.i.d. frailty model and the ICAR frailty model. The CAR model presented the best performances in presence of spatial correlation, thus demonstrating a good quality of adjustment of this model on the latter. In a last simulation study, we have shown that the performance of the CAR model remains correct when the percentage of censored observations does not exceed 20\%.

These approaches were then applied to epidemiological data to detect the presence of clusters of abnormally low or high survival times in ESRD elderly patients in northern France during the period from 2004 to 2020.
The classical approaches \citep{cook2007spatial, huang2007spatial} detected two statistically significant clusters: one in the northeast of the region corresponding to higher survival times (lower risk) and the other containing the whole western part of the region corresponding to lower survival times (higher risk) than elsewhere. The i.i.d. shared frailty model detected only as statistically significant the cluster in the northeast of the region. Assuming a complete spatial correlation (ICAR model), the method also identified a MLC in the northeast of the region but this latter was not statistically significant. When we considered the CAR frailty model allowing a flexibility of the spatial correlation, a statistically significant cluster was detected in the northeast of the region with a p-value slightly higher than the one provided by the i.i.d. shared frailty model. This can be explained by the consideration by the CAR model of the spatial correlation. These results are consistent with those of the simulation study.

In both the simulation study and the application, circular potential clusters have been considered. However, other cluster shapes (e.g., elliptical clusters \citep{elliptic}) could be considered, as the shape of the scanning window has an impact on the power of the method in cluster detection. It should be noted that the choice of another form of scanning window can be easily implemented in our method because it only changes the set of potential clusters $\mathcal{W}$. 

In the application on epidemiological data of elderly patients with ESRD, it was observed a low percentage of patients who received a renal transplant. However, this percentage is higher in the general population \citep{couchoud2015development}. It is well known that renal transplantation is a competing risk of death in patients with ESRD, and its non-consideration in the analysis may lead to a biased estimate of the survival function \citep{Hallan_2012}. In this context, the proposed method should be modified to account for competing risks by considering, for example, the model proposed by \cite{fine1999proportional}.

Here the spatial correlation parameter $\rho$ was assumed to be constant over the whole study area. This assumption may be too simplistic because this coefficient may vary spatially \citep{CRAWFORD200929}. The integration of a spatial correlation coefficient that can vary spatially appears challenging because it is necessary to clearly distinguish the effect of spatial correlation from the effects of spatial clusters present in the data. Adapting the proposed method to this context may be the subject of future work.

In our model, we have included the covariates as fixed effects. However, it is possible to consider them as random effects that may have a spatial correlation. One way to take into account these spatially structured effects is to use conditional autoregressive models for the prior distributions of the coefficients associated with the covariates.

Lastly, our spatial scan statistic can be extended to deal with recurrent events. For example, one may be interested in the time until asthma attack in patients treated for asthma: a patient may experience several asthma attacks during the study period. One possible approach is to consider a shared frailty at the individual level, making it possible to take into account unobserved subject-specific factors \citep{kleinbaum2012recurrent}. However, the time to asthma attack may also exhibit an intra-spatial unit correlation (due to environmental factors for example). In this context, one approach would be to consider a nested frailty model \citep{rondeau2010statistical}, i.e., a model with both shared frailties at the level of spatial units with a potential spatial correlation, and shared frailties at the level of individuals, in order to take into account both unobserved factors specific to spatial units (e.g., air quality), and unobserved factors specific to individuals (e.g., their tobacco consumption).

\newpage
\bibliography{bibliographie.bib}

\begin{thebibliography}{}

\bibitem[\protect\citeauthoryear{Abolhassani and Prates}{Abolhassani and
  Prates}{2021}]{abolhassani2021up}
Abolhassani, A. and M.~O. Prates (2021).
\newblock An up-to-date review of scan statistics.
\newblock {\em Statistics Surveys\/}~{\em 15}, 111--153.

\bibitem[\protect\citeauthoryear{Ahmed, Cucala, and Genin}{Ahmed
  et~al.}{2021}]{ahmed2021spatial}
Ahmed, M.-S., L.~Cucala, and M.~Genin (2021).
\newblock Spatial autoregressive models for scan statistic.
\newblock {\em Journal of Spatial Econometrics\/}~{\em 2\/}(1), 1--20.

\bibitem[\protect\citeauthoryear{Ahmed and Genin}{Ahmed and
  Genin}{2020}]{ahmed2020functional}
Ahmed, M.-S. and M.~Genin (2020).
\newblock A functional-model-adjusted spatial scan statistic.
\newblock {\em Statistics in Medicine\/}~{\em 39\/}(8), 1025--1040.

\bibitem[\protect\citeauthoryear{Arlinghaus}{Arlinghaus}{1995}]{arlinghaus1995practical}
Arlinghaus, S. (1995).
\newblock {\em Practical handbook of spatial statistics}.
\newblock CRC press.

\bibitem[\protect\citeauthoryear{Aswi, Cramb, Duncan, Hu, White, and
  Mengersen}{Aswi et~al.}{2020}]{aswi2020bayesian}
Aswi, A., S.~Cramb, E.~Duncan, W.~Hu, G.~White, and K.~Mengersen (2020).
\newblock Bayesian spatial survival models for hospitalisation of dengue: A
  case study of wahidin hospital in makassar, indonesia.
\newblock {\em International journal of environmental research and public
  health\/}~{\em 17\/}(3), 878.

\bibitem[\protect\citeauthoryear{Austin}{Austin}{2017}]{austin2017tutorial}
Austin, P.~C. (2017).
\newblock A tutorial on multilevel survival analysis: methods, models and
  applications.
\newblock {\em International Statistical Review\/}~{\em 85\/}(2), 185--203.

\bibitem[\protect\citeauthoryear{Ayav, Beuscart, Brian{\c{c}}on, Duhamel,
  Frimat, and Kessler}{Ayav et~al.}{2016}]{ayav2016competing}
Ayav, C., J.-B. Beuscart, S.~Brian{\c{c}}on, A.~Duhamel, L.~Frimat, and
  M.~Kessler (2016).
\newblock Competing risk of death and end-stage renal disease in incident
  chronic kidney disease (stages 3 to 5): the epiran community-based study.
\newblock {\em BMC nephrology\/}~{\em 17\/}(1), 1--13.

\bibitem[\protect\citeauthoryear{Banerjee, Wall, and Carlin}{Banerjee
  et~al.}{2003}]{banerjee2003frailty}
Banerjee, S., M.~M. Wall, and B.~P. Carlin (2003).
\newblock Frailty modeling for spatially correlated survival data, with
  application to infant mortality in minnesota.
\newblock {\em Biostatistics\/}~{\em 4\/}(1), 123--142.

\bibitem[\protect\citeauthoryear{Besag}{Besag}{1974}]{besag}
Besag, J. (1974).
\newblock Spatial interaction and the statistical analysis of lattice systems.
\newblock {\em Journal of the Royal Statistical Society. Series B
  (Methodological)\/}~{\em 36\/}(2), 192–236.

\bibitem[\protect\citeauthoryear{Besag, York, and Molli{\'e}}{Besag
  et~al.}{1991}]{besag1991bayesian}
Besag, J., J.~York, and A.~Molli{\'e} (1991).
\newblock Bayesian image restoration, with two applications in spatial
  statistics.
\newblock {\em Annals of the institute of statistical mathematics\/}~{\em
  43\/}(1), 1--20.

\bibitem[\protect\citeauthoryear{Bhatt and Tiwari}{Bhatt and
  Tiwari}{2014}]{bhatt2014spatial}
Bhatt, V. and N.~Tiwari (2014).
\newblock A spatial scan statistic for survival data based on weibull
  distribution.
\newblock {\em Statistics in medicine\/}~{\em 33\/}(11), 1867--1876.

\bibitem[\protect\citeauthoryear{Brook}{Brook}{1964}]{Brook}
Brook, D. (1964).
\newblock On the distinction between the conditional probability and the joint
  probability approaches in the specification of nearest-neighbour systems.
\newblock {\em Biometrika\/}~{\em 51}, 481–483.

\bibitem[\protect\citeauthoryear{Can{\c{c}}ado, da~Silva, and
  da~Silva}{Can{\c{c}}ado et~al.}{2014}]{canccado2014spatial}
Can{\c{c}}ado, A.~L., C.~Q. da~Silva, and M.~F. da~Silva (2014).
\newblock A spatial scan statistic for zero-inflated poisson process.
\newblock {\em Environmental and ecological statistics\/}~{\em 21\/}(4),
  627--650.

\bibitem[\protect\citeauthoryear{Can{\c{c}}ado, Fernandes, and
  da~Silva}{Can{\c{c}}ado et~al.}{2017}]{canccado2017bayesian}
Can{\c{c}}ado, A.~L., L.~B. Fernandes, and C.~Q. da~Silva (2017).
\newblock A bayesian spatial scan statistic for zero-inflated count data.
\newblock {\em Spatial Statistics\/}~{\em 20}, 57--75.

\bibitem[\protect\citeauthoryear{Chen and Glaz}{Chen and
  Glaz}{2009}]{rectangular}
Chen, J. and J.~Glaz (2009).
\newblock Approximations for two-dimensional variable window scan statistics.
\newblock In J.~Glaz, V.~Pozdnyakov, and S.~Wallenstein (Eds.), {\em Scan
  Statistics}, pp.\  109 -- 128. Birkh{\"a}user Boston.

\bibitem[\protect\citeauthoryear{Clayton}{Clayton}{1978}]{clayton1978model}
Clayton, D.~G. (1978).
\newblock A model for association in bivariate life tables and its application
  in epidemiological studies of familial tendency in chronic disease incidence.
\newblock {\em Biometrika\/}~{\em 65\/}(1), 141--151.

\bibitem[\protect\citeauthoryear{Cook, Gold, and Li}{Cook
  et~al.}{2007}]{cook2007spatial}
Cook, A.~J., D.~R. Gold, and Y.~Li (2007).
\newblock Spatial cluster detection for censored outcome data.
\newblock {\em Biometrics\/}~{\em 63\/}(2), 540--549.

\bibitem[\protect\citeauthoryear{Couchoud, Stengel, Landais, Aldigier,
  de~Cornelissen, Dabot, Maheut, Joyeux, Kessler, Labeeuw, Isnard, and
  Jacquelinet}{Couchoud et~al.}{2005}]{Couchoud_2005}
Couchoud, C., B.~Stengel, P.~Landais, J.-C. Aldigier, F.~de~Cornelissen,
  C.~Dabot, H.~Maheut, V.~Joyeux, M.~Kessler, M.~Labeeuw, H.~Isnard, and
  C.~Jacquelinet (2005, 10).
\newblock {The renal epidemiology and information network (REIN): a new
  registry for end-stage renal disease in France}.
\newblock {\em Nephrology Dialysis Transplantation\/}~{\em 21\/}(2), 411--418.

\bibitem[\protect\citeauthoryear{Couchoud, Beuscart, Aldigier, Brunet, and
  Moranne}{Couchoud et~al.}{2015}]{couchoud2015development}
Couchoud, C.~G., J.-B.~R. Beuscart, J.-C. Aldigier, P.~J. Brunet, and O.~P.
  Moranne (2015).
\newblock Development of a risk stratification algorithm to improve
  patient-centered care and decision making for incident elderly patients with
  end-stage renal disease.
\newblock {\em Kidney international\/}~{\em 88\/}(5), 1178--1186.

\bibitem[\protect\citeauthoryear{Crawford}{Crawford}{2009}]{CRAWFORD200929}
Crawford, T. (2009).
\newblock Scale analytical.
\newblock In R.~Kitchin and N.~Thrift (Eds.), {\em International Encyclopedia
  of Human Geography}, pp.\  29--36. Oxford: Elsevier.

\bibitem[\protect\citeauthoryear{Cressie}{Cressie}{1977}]{cressie}
Cressie, N. (1977).
\newblock On some properties of the scan statistic on the circle and the line.
\newblock {\em Journal of Applied Probability\/}~{\em 14}, 272--283.

\bibitem[\protect\citeauthoryear{Cucala, Genin, Lanier, and Occelli}{Cucala
  et~al.}{2017}]{a_multivariate_gaussian}
Cucala, L., M.~Genin, C.~Lanier, and F.~Occelli (2017).
\newblock A multivariate gaussian scan statistic for spatial data.
\newblock {\em Spatial Statistics\/}~{\em 21}.

\bibitem[\protect\citeauthoryear{De~La Fuente~Marcos and De~La
  Fuente~Marcos}{De~La Fuente~Marcos and De~La
  Fuente~Marcos}{2008}]{de2008star}
De~La Fuente~Marcos, R. and C.~De~La Fuente~Marcos (2008).
\newblock From star complexes to the field: open cluster families.
\newblock {\em The Astrophysical Journal\/}~{\em 672\/}(1), 342.

\bibitem[\protect\citeauthoryear{de~Lima, Duczmal, Neto, and Pinto}{de~Lima
  et~al.}{2015}]{de2015spatial}
de~Lima, M.~S., L.~H. Duczmal, J.~C. Neto, and L.~P. Pinto (2015).
\newblock Spatial scan statistics for models with overdispersion and inflated
  zeros.
\newblock {\em Statistica Sinica\/}, 225--241.

\bibitem[\protect\citeauthoryear{Dwass}{Dwass}{1957}]{dwass}
Dwass, M. (1957).
\newblock Modified randomization tests for nonparametric hypotheses.
\newblock {\em Annals of Mathematical Statistics\/}~{\em 28\/}(1), 181--187.

\bibitem[\protect\citeauthoryear{Fine and Gray}{Fine and
  Gray}{1999}]{fine1999proportional}
Fine, J.~P. and R.~J. Gray (1999).
\newblock A proportional hazards model for the subdistribution of a competing
  risk.
\newblock {\em Journal of the American statistical association\/}~{\em
  94\/}(446), 496--509.

\bibitem[\protect\citeauthoryear{Fr{\'e}vent, Ahmed, Marbac, and
  Genin}{Fr{\'e}vent et~al.}{2021}]{frevent2021detecting}
Fr{\'e}vent, C., M.-S. Ahmed, M.~Marbac, and M.~Genin (2021, 12).
\newblock Detecting spatial clusters in functional data: new scan statistic
  approaches.
\newblock {\em Spatial Statistics\/}~{\em 46}, 100550.

\bibitem[\protect\citeauthoryear{Frévent, Ahmed, Dabo-Niang, and
  Genin}{Frévent et~al.}{2021}]{frevent2021investigating}
Frévent, C., M.-S. Ahmed, S.~Dabo-Niang, and M.~Genin (2021, 03).
\newblock Investigating spatial scan statistics for multivariate functional
  data.
\newblock {\em arXiv preprint arXiv:2103.14401\/}.

\bibitem[\protect\citeauthoryear{Fu, Evans, Carrero, Putter, Clase, Caskey,
  Szymczak, Torino, Chesnaye, Jager, et~al.}{Fu et~al.}{2021}]{fu2021timing}
Fu, E.~L., M.~Evans, J.-J. Carrero, H.~Putter, C.~M. Clase, F.~J. Caskey,
  M.~Szymczak, C.~Torino, N.~C. Chesnaye, K.~J. Jager, et~al. (2021).
\newblock Timing of dialysis initiation to reduce mortality and cardiovascular
  events in advanced chronic kidney disease: nationwide cohort study.
\newblock {\em bmj\/}~{\em 375}.

\bibitem[\protect\citeauthoryear{Genin, Fumery, Occelli, Savoye, Pariente,
  Dauchet, Giovannelli, Vignal, Body-Malapel, Sarter, et~al.}{Genin
  et~al.}{2020}]{genin2020fine}
Genin, M., M.~Fumery, F.~Occelli, G.~Savoye, B.~Pariente, L.~Dauchet,
  J.~Giovannelli, C.~Vignal, M.~Body-Malapel, H.~Sarter, et~al. (2020).
\newblock Fine-scale geographical distribution and ecological risk factors for
  crohn's disease in france (2007-2014).
\newblock {\em Alimentary Pharmacology \& Therapeutics\/}~{\em 51\/}(1),
  139--148.

\bibitem[\protect\citeauthoryear{Green, Elliott, Beaudoin, and Bernstein}{Green
  et~al.}{2006}]{green2006population}
Green, C., L.~Elliott, C.~Beaudoin, and C.~N. Bernstein (2006).
\newblock A population-based ecologic study of inflammatory bowel disease:
  searching for etiologic clues.
\newblock {\em American journal of epidemiology\/}~{\em 164\/}(7), 615--623.

\bibitem[\protect\citeauthoryear{Gregorio, Huang, DeChello, Samociuk, and
  Kulldorff}{Gregorio et~al.}{2007}]{gregorio2007place}
Gregorio, D.~I., L.~Huang, L.~M. DeChello, H.~Samociuk, and M.~Kulldorff
  (2007).
\newblock Place of residence effect on likelihood of surviving prostate cancer.
\newblock {\em Annals of epidemiology\/}~{\em 17\/}(7), 520--524.

\bibitem[\protect\citeauthoryear{Hallan, Matsushita, Sang, Mahmoodi, Black,
  Ishani, Kleefstra, Naimark, Roderick, Tonelli, Wetzels, Astor, Gansevoort,
  Levin, Wen, Coresh, and Chronic Kidney Disease Prognosis~Consortium}{Hallan
  et~al.}{2012}]{Hallan_2012}
Hallan, S.~I., K.~Matsushita, Y.~Sang, B.~K. Mahmoodi, C.~Black, A.~Ishani,
  N.~Kleefstra, D.~Naimark, P.~Roderick, M.~Tonelli, J.~F.~M. Wetzels, B.~C.
  Astor, R.~T. Gansevoort, A.~Levin, C.-P. Wen, J.~Coresh, and f.~t. Chronic
  Kidney Disease Prognosis~Consortium (2012, 12).
\newblock {Age and Association of Kidney Measures With Mortality and End-stage
  Renal Disease}.
\newblock {\em JAMA\/}~{\em 308\/}(22), 2349--2360.

\bibitem[\protect\citeauthoryear{Henry, Niu, and Boscoe}{Henry
  et~al.}{2009}]{henry2009geographic}
Henry, K.~A., X.~Niu, and F.~P. Boscoe (2009).
\newblock Geographic disparities in colorectal cancer survival.
\newblock {\em International journal of health geographics\/}~{\em 8\/}(1),
  1--13.

\bibitem[\protect\citeauthoryear{Hougaard}{Hougaard}{2000}]{Hougaard2000}
Hougaard, P. (2000).
\newblock {\em Shared frailty models}, pp.\  215--262.
\newblock New York, NY: Springer New York.

\bibitem[\protect\citeauthoryear{Huang, Kulldorff, and Gregorio}{Huang
  et~al.}{2007}]{huang2007spatial}
Huang, L., M.~Kulldorff, and D.~Gregorio (2007).
\newblock A spatial scan statistic for survival data.
\newblock {\em Biometrics\/}~{\em 63\/}(1), 109--118.

\bibitem[\protect\citeauthoryear{Jeffreys}{Jeffreys}{1961}]{jeffreys1998theory}
Jeffreys, H. (1961).
\newblock {\em Theory of probability (3rd Ed.)}.
\newblock Oxford, UK: Oxford University Press.

\bibitem[\protect\citeauthoryear{Jung}{Jung}{2009}]{jung2009generalized}
Jung, I. (2009).
\newblock A generalized linear models approach to spatial scan statistics for
  covariate adjustment.
\newblock {\em Statistics in medicine\/}~{\em 28\/}(7), 1131--1143.

\bibitem[\protect\citeauthoryear{Jung, Kulldorff, and Klassen}{Jung
  et~al.}{2007}]{jung2007spatial}
Jung, I., M.~Kulldorff, and A.~C. Klassen (2007).
\newblock A spatial scan statistic for ordinal data.
\newblock {\em Statistics in medicine\/}~{\em 26\/}(7), 1594--1607.

\bibitem[\protect\citeauthoryear{Khan, Roberson, Reid, Jordan, and Odoi}{Khan
  et~al.}{2021}]{khan2021geographic}
Khan, M.~M., S.~Roberson, K.~Reid, M.~Jordan, and A.~Odoi (2021).
\newblock Geographic disparities and temporal changes of diabetes prevalence
  and diabetes self-management education program participation in florida.
\newblock {\em Plos one\/}~{\em 16\/}(7).

\bibitem[\protect\citeauthoryear{Kleinbaum and Klein}{Kleinbaum and
  Klein}{2012}]{kleinbaum2012recurrent}
Kleinbaum, D.~G. and M.~Klein (2012).
\newblock Recurrent event survival analysis.
\newblock In {\em Survival Analysis}, pp.\  363--423. Springer.

\bibitem[\protect\citeauthoryear{Kulldorff}{Kulldorff}{1997}]{kulldorff1997spatial}
Kulldorff, M. (1997).
\newblock A spatial scan statistic.
\newblock {\em Communications in Statistics - Theory and Methods\/}~{\em 26},
  1481--1496.

\bibitem[\protect\citeauthoryear{Kulldorff, Huang, and Konty}{Kulldorff
  et~al.}{2009}]{normalkulldorff}
Kulldorff, M., L.~Huang, and K.~Konty (2009).
\newblock A scan statistic for continuous data based on the normal probability
  model.
\newblock {\em Int J Health Geogr\/}~{\em 8\/}(58).

\bibitem[\protect\citeauthoryear{Kulldorff, Huang, Pickle, and
  Duczmal}{Kulldorff et~al.}{2006}]{elliptic}
Kulldorff, M., L.~Huang, L.~Pickle, and L.~Duczmal (2006).
\newblock An elliptic spatial scan statistic.
\newblock {\em Statistics in medicine\/}~{\em 25}, 3929--3943.

\bibitem[\protect\citeauthoryear{Kulldorff, Mostashari, Duczmal, Katherine~Yih,
  Kleinman, and Platt}{Kulldorff et~al.}{2007}]{kulldorffmulti}
Kulldorff, M., F.~Mostashari, L.~Duczmal, W.~Katherine~Yih, K.~Kleinman, and
  R.~Platt (2007).
\newblock Multivariate scan statistics for disease surveillance.
\newblock {\em Statistics in medicine\/}~{\em 26\/}(8), 1824--1833.

\bibitem[\protect\citeauthoryear{Kulldorff and Nagarwalla}{Kulldorff and
  Nagarwalla}{1995}]{spatialdisease}
Kulldorff, M. and N.~Nagarwalla (1995).
\newblock Spatial disease clusters: Detection and inference.
\newblock {\em Statistics in Medicine\/}~{\em 14\/}(8), 799--810.

\bibitem[\protect\citeauthoryear{Lee, Sun, and Chang}{Lee
  et~al.}{2020}]{lee2020spatial}
Lee, J., Y.~Sun, and H.~H. Chang (2020).
\newblock Spatial cluster detection of regression coefficients in a
  mixed-effects model.
\newblock {\em Environmetrics\/}~{\em 31\/}(2), e2578.

\bibitem[\protect\citeauthoryear{Leiser, Taddie, Hemmert, Richards~Steed,
  VanDerslice, Henry, Ambrose, O’Neil, Smith, and Hanson}{Leiser
  et~al.}{2020}]{leiser2020spatial}
Leiser, C.~L., M.~Taddie, R.~Hemmert, R.~Richards~Steed, J.~A. VanDerslice,
  K.~Henry, J.~Ambrose, B.~O’Neil, K.~R. Smith, and H.~A. Hanson (2020).
\newblock Spatial clusters of cancer incidence: Analyzing 1940 census data
  linked to 1966--2017 cancer records.
\newblock {\em Cancer Causes \& Control\/}~{\em 31\/}(7), 609--615.

\bibitem[\protect\citeauthoryear{Leroux, Lei, and Breslow}{Leroux
  et~al.}{2000}]{leroux}
Leroux, B.~G., X.~Lei, and N.~Breslow (2000).
\newblock Estimation of disease rates in small areas: a new mixed model for
  spatial dependence.
\newblock In {\em Statistical models in epidemiology, the environment, and
  clinical trials}, pp.\  179--191. Springer.

\bibitem[\protect\citeauthoryear{Li}{Li}{2009}]{li2009modeling}
Li, Y. (2009).
\newblock Modeling and analysis of spatially correlated data.
\newblock {\em New Developments In Biostatistics And Bioinformatics\/}, 72--98.

\bibitem[\protect\citeauthoryear{Li and Ryan}{Li and
  Ryan}{2002}]{li2002modeling}
Li, Y. and L.~Ryan (2002).
\newblock Modeling spatial survival data using semiparametric frailty models.
\newblock {\em Biometrics\/}~{\em 58\/}(2), 287--297.

\bibitem[\protect\citeauthoryear{Liang, Self, Bandeen-Roche, and Zeger}{Liang
  et~al.}{1995}]{liang1995some}
Liang, K.-Y., S.~G. Self, K.~J. Bandeen-Roche, and S.~L. Zeger (1995).
\newblock Some recent developments for regression analysis of multivariate
  failure time data.
\newblock {\em Lifetime data analysis\/}~{\em 1\/}(4), 403--415.

\bibitem[\protect\citeauthoryear{Lin}{Lin}{2014}]{lin2014generalized}
Lin, P.-S. (2014).
\newblock Generalized scan statistics for disease surveillance.
\newblock {\em Scandinavian Journal of Statistics\/}~{\em 41\/}(3), 791--808.

\bibitem[\protect\citeauthoryear{Loh and Zhu}{Loh and
  Zhu}{2007}]{loh2007accounting}
Loh, J.~M. and Z.~Zhu (2007).
\newblock Accounting for spatial correlation in the scan statistic.
\newblock {\em The Annals of Applied Statistics\/}~{\em 1\/}(2), 560--584.

\bibitem[\protect\citeauthoryear{Marciano, de~Faria Fernandes~Belone, Rosa,
  Coelho, Ghidella, Nardi, Miranda, Barrozo, and Last{\'o}ria}{Marciano
  et~al.}{2018}]{marciano2018epidemiological}
Marciano, L. H. S.~C., A.~de~Faria Fernandes~Belone, P.~S. Rosa, N.~M.~B.
  Coelho, C.~C. Ghidella, S.~M.~T. Nardi, W.~C. Miranda, L.~V. Barrozo, and
  J.~C. Last{\'o}ria (2018).
\newblock Epidemiological and geographical characterization of leprosy in a
  brazilian hyperendemic municipality.
\newblock {\em Cadernos de saude publica\/}~{\em 34}.

\bibitem[\protect\citeauthoryear{Minamisava, Nouer, de~Morais~Neto, Melo, and
  Andrade}{Minamisava et~al.}{2009}]{minamisava2009spatial}
Minamisava, R., S.~S. Nouer, O.~L. de~Morais~Neto, L.~K. Melo, and A.~L.~S.
  Andrade (2009).
\newblock Spatial clusters of violent deaths in a newly urbanized region of
  brazil: highlighting the social disparities.
\newblock {\em International journal of health geographics\/}~{\em 8\/}(1),
  1--10.

\bibitem[\protect\citeauthoryear{Montez-Rath, Kapphahn, Mathur, Mitani, Hendry,
  and Desai}{Montez-Rath et~al.}{2017}]{montez2017guidelines}
Montez-Rath, M.~E., K.~Kapphahn, M.~B. Mathur, A.~A. Mitani, D.~J. Hendry, and
  M.~Desai (2017).
\newblock Guidelines for generating right-censored outcomes from a cox model
  extended to accommodate time-varying covariates.
\newblock {\em Journal of Modern Applied Statistical Methods\/}~{\em 16\/}(1),
  6.

\bibitem[\protect\citeauthoryear{Neill and Cooper}{Neill and
  Cooper}{2010}]{neill2010multivariate}
Neill, D.~B. and G.~F. Cooper (2010).
\newblock A multivariate bayesian scan statistic for early event detection and
  characterization.
\newblock {\em Machine learning\/}~{\em 79\/}(3), 261--282.

\bibitem[\protect\citeauthoryear{Ojiambo and Kang}{Ojiambo and
  Kang}{2013}]{ojiambo2013modeling}
Ojiambo, P. and E.~Kang (2013).
\newblock Modeling spatial frailties in survival analysis of cucurbit downy
  mildew epidemics.
\newblock {\em Phytopathology\/}~{\em 103\/}(3), 216--227.

\bibitem[\protect\citeauthoryear{Rondeau}{Rondeau}{2010}]{rondeau2010statistical}
Rondeau, V. (2010).
\newblock Statistical models for recurrent events and death: Application to
  cancer events.
\newblock {\em Mathematical and Computer modelling\/}~{\em 52\/}(7-8),
  949--955.

\bibitem[\protect\citeauthoryear{Rue, Martino, and Chopin}{Rue
  et~al.}{2009}]{rue2009approximate}
Rue, H., S.~Martino, and N.~Chopin (2009).
\newblock Approximate bayesian inference for latent gaussian models by using
  integrated nested laplace approximations.
\newblock {\em Journal of the royal statistical society: Series b (statistical
  methodology)\/}~{\em 71\/}(2), 319--392.

\bibitem[\protect\citeauthoryear{Shi, Liu, and Zhong}{Shi
  et~al.}{2021}]{shi2021spatial}
Shi, G., J.~Liu, and X.~Zhong (2021).
\newblock Spatial and temporal variations of pm2.5 concentrations in chinese
  cities during 2015-2019.
\newblock {\em International Journal of Environmental Health Research\/},
  1--13.

\bibitem[\protect\citeauthoryear{Smida, Cucala, Gannoun, and Durif}{Smida
  et~al.}{2022}]{smida2022wilcoxon}
Smida, Z., L.~Cucala, A.~Gannoun, and G.~Durif (2022).
\newblock A wilcoxon-mann-whitney spatial scan statistic for functional data.
\newblock {\em Computational Statistics \& Data Analysis\/}~{\em 167}, 107378.

\bibitem[\protect\citeauthoryear{Tango and Takahashi}{Tango and
  Takahashi}{2005}]{tango2005flexibly}
Tango, T. and K.~Takahashi (2005).
\newblock A flexibly shaped spatial scan statistic for detecting clusters.
\newblock {\em International journal of health geographics\/}~{\em 4\/}(1),
  1--15.

\bibitem[\protect\citeauthoryear{Usman and Rosychuk}{Usman and
  Rosychuk}{2018}]{usman2018log}
Usman, I. and R.~J. Rosychuk (2018).
\newblock A log-weibull spatial scan statistic for time to event data.
\newblock {\em International Journal of Health Geographics\/}~{\em 17\/}(1),
  1--12.

\bibitem[\protect\citeauthoryear{Wan, Sun, Lee, Zhao, and Xia}{Wan
  et~al.}{2020}]{wan2020industrial}
Wan, L., Y.~Sun, I.~Lee, W.~Zhao, and F.~Xia (2020).
\newblock Industrial pollution areas detection and location via satellite-based
  iiot.
\newblock {\em IEEE Transactions on Industrial Informatics\/}~{\em 17\/}(3),
  1785--1794.

\bibitem[\protect\citeauthoryear{Wan, Zhan, Lu, and Tiefenbacher}{Wan
  et~al.}{2012}]{wan2012access}
Wan, N., F.~B. Zhan, Y.~Lu, and J.~P. Tiefenbacher (2012).
\newblock Access to healthcare and disparities in colorectal cancer survival in
  texas.
\newblock {\em Health \& place\/}~{\em 18\/}(2), 321--329.

\bibitem[\protect\citeauthoryear{Yin and Mu}{Yin and Mu}{2018}]{yin2018hybrid}
Yin, P. and L.~Mu (2018).
\newblock A hybrid method for fast detection of spatial disease clusters in
  irregular shapes.
\newblock {\em GeoJournal\/}~{\em 83\/}(4), 693--705.

\bibitem[\protect\citeauthoryear{Zhang, Zhang, and Lin}{Zhang
  et~al.}{2012}]{zhang2012spatial}
Zhang, T., Z.~Zhang, and G.~Lin (2012).
\newblock Spatial scan statistics with overdispersion.
\newblock {\em Statistics in medicine\/}~{\em 31\/}(8), 762--774.

\bibitem[\protect\citeauthoryear{Zhou, Shu, and Su}{Zhou
  et~al.}{2015}]{zhou2015adaptive}
Zhou, R., L.~Shu, and Y.~Su (2015).
\newblock An adaptive minimum spanning tree test for detecting
  irregularly-shaped spatial clusters.
\newblock {\em Computational Statistics \& Data Analysis\/}~{\em 89}, 134--146.

\end{thebibliography}
\bibliographystyle{chicago}

\newpage
\appendix
\section{Leroux prior} \label{sec:leroux}
The Leroux CAR prior is defined by \\

$X_k|X_{-k} \sim \mathcal{N}\left(\dfrac{\rho \dsum_{l=1}^{K} v_{k,l} X_l}{\rho \dsum_{l=1}^{K} v_{k,l} + 1 - \rho} ; \dfrac{\sigma_X^2}{\rho \dsum_{l=1}^{K} v_{k,l} + 1 - \rho} \right)$ where $v_{k,l}$=1 if $s_k$ and $s_l$ are adjacent and 0 otherwise. \\
Let us show (\cite{besag}) that this is equivalent to $\bm{X} \sim \mathcal{N}(0, \sigma^2_{X}[\rho R + (1-\rho)I_K]^{-1})$, where $R_{k,l} = \left\{ \begin{array}{ll}
\dsum_{j=1}^K v_{k,j} & \text{ if } k = l \\
- v_{k,l} & \text{ otherwise}
\end{array}  \right.$, \\

by applying Brook's lemma (\cite{Brook}) on $\bm{X}$: 
\begin{align*}
\dfrac{\mathbb{P}(\bm{X})}{\mathbb{P}(\bm{0})} &= \dprod_{k=1}^K \dfrac{\mathbb{P}(X_k|X_1, \dots X_{k-1}, 0_{k+1}, \dots, 0_K)}{\mathbb{P}(0_k | X_1, \dots X_{k-1}, 0_{k+1}, \dots, 0_K)}.
\end{align*}
Let $a_k$ and $b_{k,l}$ be defined as $a_k = \dfrac{\sigma_X^2}{\rho \dsum_{j=1}^K v_{k,j} + 1 - \rho}$ and $b_{k,l} = \dfrac{v_{k,l}}{\rho \dsum_{j=1}^K v_{k,j} + 1 - \rho}$, then \\

\begin{align*}
\dfrac{\mathbb{P}(\bm{X})}{\mathbb{P}(\bm{0})} &= \dprod_{k=1}^K \dfrac{\exp{\left[ - \dfrac{1}{2 a_k} \left(X_k - \rho \dsum_{l=1}^{k-1} b_{k,l} X_l \right)^2 \right]}}{\exp{\left[ - \dfrac{1}{2 a_k} \left(0_k - \rho \dsum_{l=1}^{k-1} b_{k,l} X_l \right)^2 \right]}} \\
&= \dprod_{k=1}^K \exp{\left[ - \dfrac{1}{2 a_k} \left( X_k^2 - 2\rho \dsum_{l=1}^{k-1} b_{k,l} X_k X_l \right) \right]} \\
&= \dprod_{k=1}^K \exp{\left[ - \dfrac{1}{2 \sigma_X^2} (\rho R_{k,k} + 1 - \rho) \left( X_k^2 + 2\rho \dsum_{l=1}^{k-1} \dfrac{R_{k,l}}{\rho R_{k,k} + 1 - \rho} X_k X_l \right) \right]} \\
&= \dprod_{k=1}^K \exp{\left[ - \dfrac{1}{2 \sigma_X^2} \left( (\rho R_{k,k} + 1 - \rho) X_k^2 + 2\rho \dsum_{l=1}^{k-1} R_{k,l} X_k X_l \right) \right]}. \\
\end{align*}

Let $A = \rho R + (1-\rho) I_K$, then

\begin{align*}
\dfrac{\mathbb{P}(\bm{X})}{\mathbb{P}(\bm{0})} &= \dprod_{k=1}^K \exp{\left[ - \dfrac{1}{2 \sigma_X^2} \left( A_{k,k} X_k^2 + 2 \dsum_{l=1}^{k-1} A_{k,l} X_k X_l \right) \right]} \\
&= \exp{\left[ - \dfrac{1}{2 \sigma_X^2} \left( \dsum_{k=1}^K A_{k,k} X_k^2 + 2 \dsum_{k=1}^K \dsum_{l=1}^{k-1} A_{k,l} X_k X_l \right) \right]}. \\
\end{align*}

Moreover, 
\begin{align*}
2 \dsum_{k=1}^K \dsum_{l=1}^{k-1} A_{k,l} X_k X_l &= \dsum_{k=1}^K \dsum_{l=1}^{k-1} A_{k,l} X_k X_l + \dsum_{k=1}^K \dsum_{l=1}^{k-1} A_{l,k} X_k X_l \text{ because } A \text{ is symmetric} \\
&= \dsum_{k=1}^K \dsum_{l=1}^{k-1} A_{k,l} X_k X_l + \dsum_{l=1}^K \dsum_{k=l+1}^{K} A_{l,k} X_k X_l \\
&= \dsum_{k=1}^K \dsum_{l=1}^{k-1} A_{k,l} X_k X_l + \dsum_{k=1}^K \dsum_{l=k+1}^{K} A_{k,l} X_l X_k \\
&= \dsum_{k=1}^K \dsum_{\substack{l=1 \\ l \neq k}}^{K} A_{k,l} X_k X_l. \\
\end{align*}

So 
\begin{align*}
\dfrac{\mathbb{P}(\bm{X})}{\mathbb{P}(\bm{0})} &= \exp{\left[ - \dfrac{1}{2 \sigma_X^2} \left( \dsum_{k=1}^K A_{k,k} X_k^2 + \dsum_{k=1}^K \dsum_{\substack{l=1 \\ l \neq k}}^{K} A_{k,l} X_k X_l
\right) \right]} \\
&= \exp{\left[ - \dfrac{1}{2 \sigma_X^2} \dsum_{k=1}^K \dsum_{l=1}^{K} A_{k,l} X_k X_l \right]} \\
&= \exp{\left[ - \dfrac{1}{2 \sigma_X^2} \bm{X}^\top A \bm{X} \right]}. \\
\end{align*}

Then $\mathbb{P}(\bm{X}) \propto \exp{\left[ - \dfrac{1}{2 \sigma_X^2} \bm{X}^\top A \bm{X} \right]}$ and we recognize the normal distribution $\mathcal{N}(0, \sigma_X^2 A^{-1})$.

\section{Maximum likelihood estimators of $\alpha$, $\sigma^{2(0)}$, $\alpha_w$, $\alpha_{w^\mathsf{c}}$ and $\sigma^{2(w)}$} \label{appendix:proofs}

\subsection{Estimation under $\mathcal{H}_0$}
Under $\mathcal{H}_0$ the likelihood is
$$L_{\mathcal{H}_0} = \dfrac{1}{(2\pi)^{K/2} |A^{-1}|^{1/2} \sqrt{\sigma^{2K(0)}} } \exp{\left[ -\dfrac{1}{2 \sigma^{2(0)}} [\bm{\varphi}^{*} - \alpha \mathds{1} ]^\top A [\bm{\varphi}^{*} - \alpha \mathds{1}] \right]}. $$

Then the log-likelihood is defined by
\begin{align*}
\ell_{\mathcal{H}_0} &=
-\dfrac{K}{2} \ln{[2\pi]} -\dfrac{1}{2} \ln{|A^{-1}|} - \dfrac{K}{2} \ln{[\sigma^{2(0)}]} 
 -\dfrac{1}{2 \sigma^{2(0)}} [\bm{\varphi}^{*} - \alpha \mathds{1} ]^\top A [\bm{\varphi}^{*} - \alpha \mathds{1}] \\
 &=
-\dfrac{K}{2} \ln{[2\pi]} -\dfrac{1}{2} \ln{|A^{-1}|} - \dfrac{K}{2} \ln{[\sigma^{2(0)}]} 
 -\dfrac{1}{2 \sigma^{2(0)}} [\bm{\varphi}^{*\top} A \bm{\varphi}^{*} - 2 \alpha \mathds{1}^\top A \bm{\varphi}^{*} + \alpha^2 \mathds{1}^\top A \mathds{1}].
\end{align*}

\begin{align*}
\dfrac{\partial \ell_{\mathcal{H}_0}}{\partial \alpha} &= -\dfrac{1}{2 \sigma^{2(0)}} [-2\mathds{1}^\top A \bm{\varphi}^{*} + 2\alpha \mathds{1}^\top A \mathds{1}].
\end{align*}

Thus $\dfrac{\partial \ell_{\mathcal{H}_0}}{\partial \alpha} = 0 \iff \alpha \mathds{1}^\top A \mathds{1} = \mathds{1}^\top A \bm{\varphi}^{*} \iff \alpha = \dfrac{\mathds{1}^\top A \bm{\varphi}^{*}}{\mathds{1}^\top A \mathds{1}}$.

\begin{align*}
\dfrac{\partial \ell_{\mathcal{H}_0}}{\partial \sigma^{2(0)}} &= -\dfrac{K}{2 \sigma^{2(0)}} + \dfrac{1}{2 \sigma^{4(0)}} [\bm{\varphi}^{*\top} A \bm{\varphi}^{*} - 2 \alpha \mathds{1}^\top A \bm{\varphi}^{*} + \alpha^2 \mathds{1}^\top A \mathds{1}].
\end{align*}

Then $\dfrac{\partial \ell_{\mathcal{H}_0}}{\partial \sigma^{2(0)}} = 0 \iff \sigma^{2(0)} = \dfrac{1}{K} [\bm{\varphi}^{*\top} A \bm{\varphi}^{*} - 2 \alpha \mathds{1}^\top A \bm{\varphi}^{*} + \alpha^2 \mathds{1}^\top A \mathds{1}]$. \\

Finally $\hat{\alpha} = \dfrac{\mathds{1}^\top A \bm{\varphi}^{*}}{\mathds{1}^\top A \mathds{1}}$ and $\widehat{\sigma^{2(0)}} =  \dfrac{1}{K} [\bm{\varphi}^{*\top} A \bm{\varphi}^{*} - 2 \hat{\alpha} \mathds{1}^\top A \bm{\varphi}^{*} + \hat{\alpha}^2 \mathds{1}^\top A \mathds{1}] $.

\subsection{Estimation under $\mathcal{H}_1^{(w)}$}

Under $\mathcal{H}_1^{(w)}$ the likelihood is
$$ L_{\mathcal{H}_1^{(w)}} = \dfrac{1}{(2\pi)^{K/2} |A^{-1}|^{1/2} \sqrt{\sigma^{2K(w)}} } \exp{\left[ - \dfrac{1}{2\sigma^{2(w)}}[\bm{\varphi}^{*} - \alpha_w \mathds{1}_w - \alpha_{w^\mathsf{c}} \mathds{1}_{w^\mathsf{c}} ]^\top A [\bm{\varphi}^{*} - \alpha_w \mathds{1}_w - \alpha_{w^\mathsf{c}} \mathds{1}_{w^\mathsf{c}}] \right]}. $$
Thus the log-likelihood can be computed:
\begin{align*}
\ell_{\mathcal{H}_1^{(w)}} &= 
-\dfrac{K}{2} \ln{[2\pi]} - \dfrac{1}{2} \ln{|A^{-1}|} - \dfrac{K}{2} \ln{[\sigma^{2(w)}]}
 - \dfrac{1}{2\sigma^{2(w)}}[\bm{\varphi}^{*} - \alpha_w \mathds{1}_w - \alpha_{w^\mathsf{c}} \mathds{1}_{w^\mathsf{c}} ]^\top A [\bm{\varphi}^{*} - \alpha_w \mathds{1}_w - \alpha_{w^\mathsf{c}} \mathds{1}_{w^\mathsf{c}}] \\
 &= 
-\dfrac{K}{2} \ln{[2\pi]} - \dfrac{1}{2} \ln{|A^{-1}|} - \dfrac{K}{2} \ln{[\sigma^{2(w)}]} \\ &\hspace{0.4cm}
 - \dfrac{1}{2\sigma^{2(w)}}
 [\bm{\varphi}^{* \top} A \bm{\varphi}^{*} - 2 \alpha_w \mathds{1}_w^\top A \bm{\varphi}^{*} + \alpha_w^2 \mathds{1}_w^\top A \mathds{1}_w + 2 \alpha_w \alpha_{w^\mathsf{c}} \mathds{1}_w^\top A \mathds{1}_{w^\mathsf{c}} - 2 \alpha_{w^\mathsf{c}} \mathds{1}_{w^\mathsf{c}}^\top A \bm{\varphi}^{*} + \alpha_{w^\mathsf{c}}^2 \mathds{1}_{w^\mathsf{c}}^\top A \mathds{1}_{w^\mathsf{c}} ].
\end{align*}

\begin{align*}
\dfrac{\partial \ell_{\mathcal{H}_1^{(w)}}}{\partial \alpha_w} = -\dfrac{1}{2\sigma^{2(w)}} [-2 \mathds{1}_w^\top A \bm{\varphi}^{*} + 2 \alpha_w \mathds{1}_w^\top A \mathds{1}_w + 2 \alpha_{w^\mathsf{c}} \mathds{1}_w^\top A \mathds{1}_{w^\mathsf{c}}]. 
\end{align*}

Then $\dfrac{\partial \ell_{\mathcal{H}_1^{(w)}}}{\partial \alpha_w} = 0 \iff \alpha_w \mathds{1}_w^\top A \mathds{1}_w = \mathds{1}_w^\top A \bm{\varphi}^{*} - \alpha_{w^\mathsf{c}} \mathds{1}_w^\top A \mathds{1}_{w^\mathsf{c}}$.

\begin{align*}
\dfrac{\partial \ell_{\mathcal{H}_1^{(w)}}}{\partial \alpha_{w^\mathsf{c}}} = -\dfrac{1}{2\sigma^{2(w)}} [-2 \mathds{1}_{w^\mathsf{c}}^\top A \bm{\varphi}^{*} + 2 \alpha_{w^\mathsf{c}} \mathds{1}_{w^\mathsf{c}}^\top A \mathds{1}_{w^\mathsf{c}} + 2 \alpha_{w} \mathds{1}_w^\top A \mathds{1}_{w^\mathsf{c}}]. 
\end{align*}

Then $\dfrac{\partial \ell_{\mathcal{H}_1^{(w)}}}{\partial \alpha_{w^\mathsf{c}}} = 0 \iff \alpha_{w^\mathsf{c}} \mathds{1}_{w^\mathsf{c}}^\top A \mathds{1}_{w^\mathsf{c}} = \mathds{1}_{w^\mathsf{c}}^\top A \bm{\varphi}^{*} - \alpha_{w} \mathds{1}_w^\top A \mathds{1}_{w^\mathsf{c}}$. \\

We deduce:
\begin{align*}
\left\{ 
\begin{array}{rl}
\dfrac{\partial \ell_{\mathcal{H}_1^{(w)}}}{\partial \alpha_{w}} & = 0 \\
\dfrac{\partial \ell_{\mathcal{H}_1^{(w)}}}{\partial \alpha_{w^\mathsf{c}}} & = 0
\end{array}
\right. &\iff \left\{
\begin{array}{rl}
\alpha_w \mathds{1}_w^\top A \mathds{1}_w & = \mathds{1}_w^\top A \bm{\varphi}^{*} - \alpha_{w^\mathsf{c}} \mathds{1}_w^\top A \mathds{1}_{w^\mathsf{c}} \\
\alpha_{w^\mathsf{c}} \mathds{1}_{w^\mathsf{c}}^\top A \mathds{1}_{w^\mathsf{c}} & = \mathds{1}_{w^\mathsf{c}}^\top A \bm{\varphi}^{*} - \alpha_{w} \mathds{1}_w^\top A \mathds{1}_{w^\mathsf{c}}
\end{array}
\right. \\
&\iff \left\{
\begin{array}{rl}
\alpha_w & = \dfrac{\mathds{1}_w^\top A \bm{\varphi}^{*} - \alpha_{w^\mathsf{c}} \mathds{1}_w^\top A \mathds{1}_{w^\mathsf{c}}}{\mathds{1}_w^\top A \mathds{1}_w}  \\
\alpha_{w^\mathsf{c}} \mathds{1}_{w^\mathsf{c}}^\top A \mathds{1}_{w^\mathsf{c}} & = \mathds{1}_{w^\mathsf{c}}^\top A \bm{\varphi}^{*} - \dfrac{\mathds{1}_w^\top A \bm{\varphi}^{*} - \alpha_{w^\mathsf{c}} \mathds{1}_w^\top A \mathds{1}_{w^\mathsf{c}}}{\mathds{1}_w^\top A \mathds{1}_w} \mathds{1}_w^\top A \mathds{1}_{w^\mathsf{c}}
\end{array}
\right. 
\\
&\iff \left\{
\begin{array}{rl}
\alpha_w & = \dfrac{\mathds{1}_w^\top A \bm{\varphi}^{*} - \alpha_{w^\mathsf{c}} \mathds{1}_w^\top A \mathds{1}_{w^\mathsf{c}}}{\mathds{1}_w^\top A \mathds{1}_w}  \\
\alpha_{w^\mathsf{c}} & = \left[ \mathds{1}_{w^\mathsf{c}}^\top A \mathds{1}_{w^\mathsf{c}} - \dfrac{\mathds{1}_w^\top A \mathds{1}_{w^\mathsf{c}} \mathds{1}_w^\top A \mathds{1}_{w^\mathsf{c}} }{\mathds{1}_w^\top A \mathds{1}_w}\right]^{-1} \left[\mathds{1}_{w^\mathsf{c}}^\top A \bm{\varphi}^{*} - \dfrac{\mathds{1}_w^\top A \bm{\varphi}^{*} \mathds{1}_w^\top A \mathds{1}_{w^\mathsf{c}} }{\mathds{1}_w^\top A \mathds{1}_w} \right].
\end{array}
\right.
\end{align*}

\begin{align*}
\dfrac{\partial \ell_{\mathcal{H}_1^{(w)}}}{\partial \sigma^{2(w)}} = -\dfrac{K}{2 \sigma^{2(w)}} + \dfrac{1}{2 \sigma^{4(w)}} [\bm{\varphi}^{*} - \alpha_w \mathds{1}_w - \alpha_{w^\mathsf{c}} \mathds{1}_{w^\mathsf{c}} ]^\top A [\bm{\varphi}^{*} - \alpha_w \mathds{1}_w - \alpha_{w^\mathsf{c}} \mathds{1}_{w^\mathsf{c}}].
\end{align*}

Then $\dfrac{\partial \ell_{\mathcal{H}_1^{(w)}}}{\partial \sigma^{2(w)}} = 0 \iff \sigma^{2(w)} = \dfrac{1}{K} [\bm{\varphi}^{*} - \alpha_w \mathds{1}_w - \alpha_{w^\mathsf{c}} \mathds{1}_{w^\mathsf{c}} ]^\top A [\bm{\varphi}^{*} - \alpha_w \mathds{1}_w - \alpha_{w^\mathsf{c}} \mathds{1}_{w^\mathsf{c}}]. $ \\

Finally $\hat{\alpha}_{w^\mathsf{c}} = \left[ \mathds{1}_{w^\mathsf{c}}^\top A \mathds{1}_{w^\mathsf{c}} - \dfrac{\mathds{1}_w^\top A \mathds{1}_{w^\mathsf{c}} \mathds{1}_w^\top A \mathds{1}_{w^\mathsf{c}} }{\mathds{1}_w^\top A \mathds{1}_w}\right]^{-1} \left[\mathds{1}_{w^\mathsf{c}}^\top A \bm{\varphi}^{*} - \dfrac{\mathds{1}_w^\top A \bm{\varphi}^{*} \mathds{1}_w^\top A \mathds{1}_{w^\mathsf{c}} }{\mathds{1}_w^\top A \mathds{1}_w} \right]$, \\ $\hat{\alpha}_w = \dfrac{\mathds{1}_w^\top A \bm{\varphi}^{*} - \hat{\alpha}_{w^\mathsf{c}} \mathds{1}_w^\top A \mathds{1}_{w^\mathsf{c}}}{\mathds{1}_w^\top A \mathds{1}_w}$ and $\widehat{\sigma^{2(w)}} = \dfrac{1}{K} [\bm{\varphi}^{*} - \hat{\alpha}_w \mathds{1}_w - \hat{\alpha}_{w^\mathsf{c}} \mathds{1}_{w^\mathsf{c}} ]^\top A [\bm{\varphi}^{*} - \hat{\alpha}_w \mathds{1}_w - \hat{\alpha}_{w^\mathsf{c}} \mathds{1}_{w^\mathsf{c}}] $.

\section{Simulation study} 

\subsection{Design of the simulation study} \label{appendix:design}

\begin{figure}[h!]
    \centering
    \includegraphics[width = 10cm]{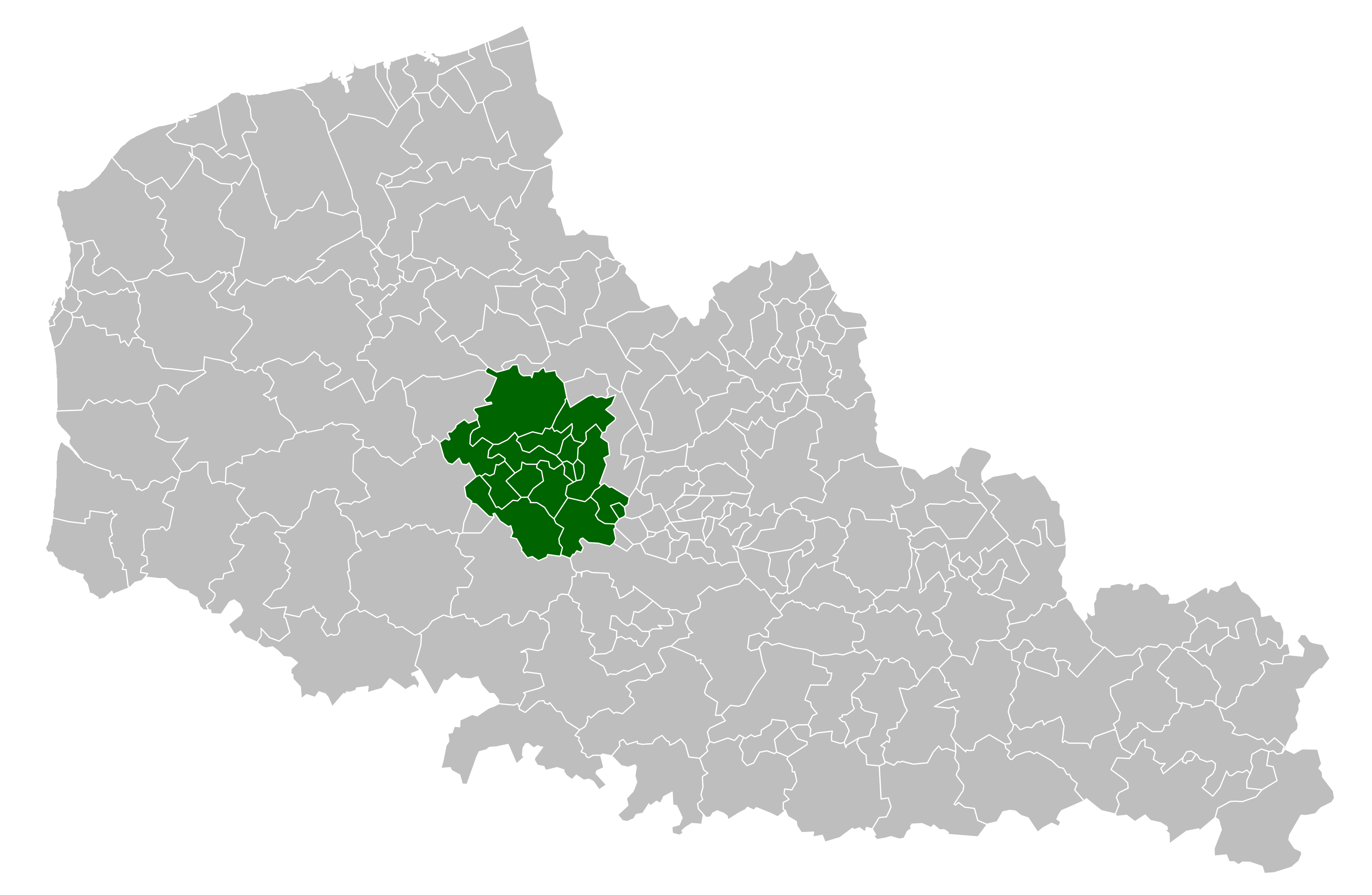}
    \caption{Simulated cluster (in green) in 169 administrative subdivisions of northern France.}
    \label{fig:map_cluster}
\end{figure}

\begin{figure}[h!]
    \centering
    \includegraphics[width = 7cm]{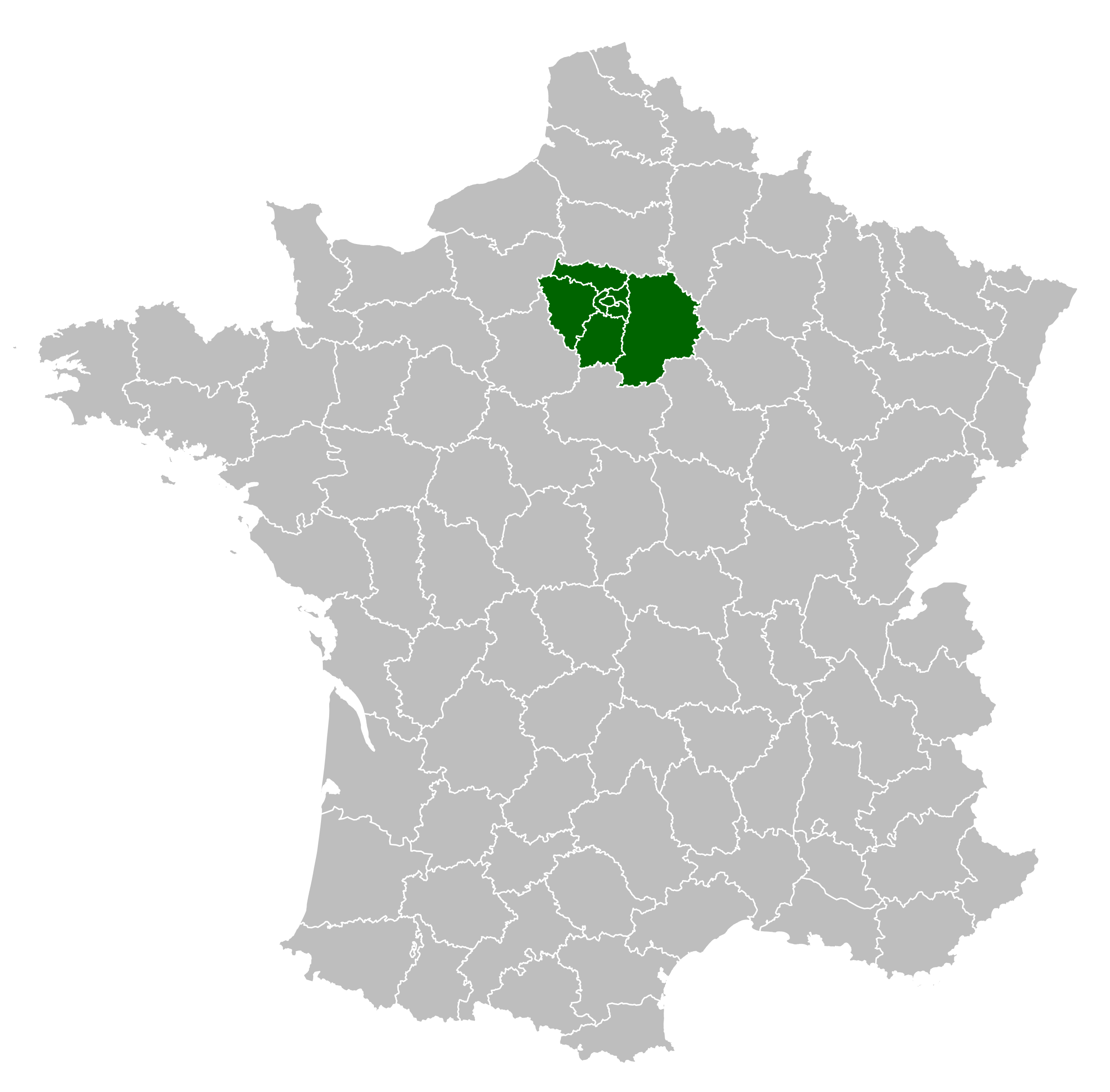}
    \caption{Simulated cluster (in green) in the 94 \textit{départements} of France.}
    \label{fig:map_cluster94}
\end{figure}

Figures \ref{fig:map_cluster} and \ref{fig:map_cluster94} show the spatial units as well as the cluster considered in the simulation study in the absence and in the presence of censoring respectively.

\subsection{Influence of the threshold chosen for the Bayes Factor} \label{appendix:otherbf}

\begin{figure}[h!]
    \centering
    \includegraphics[width=\textwidth]{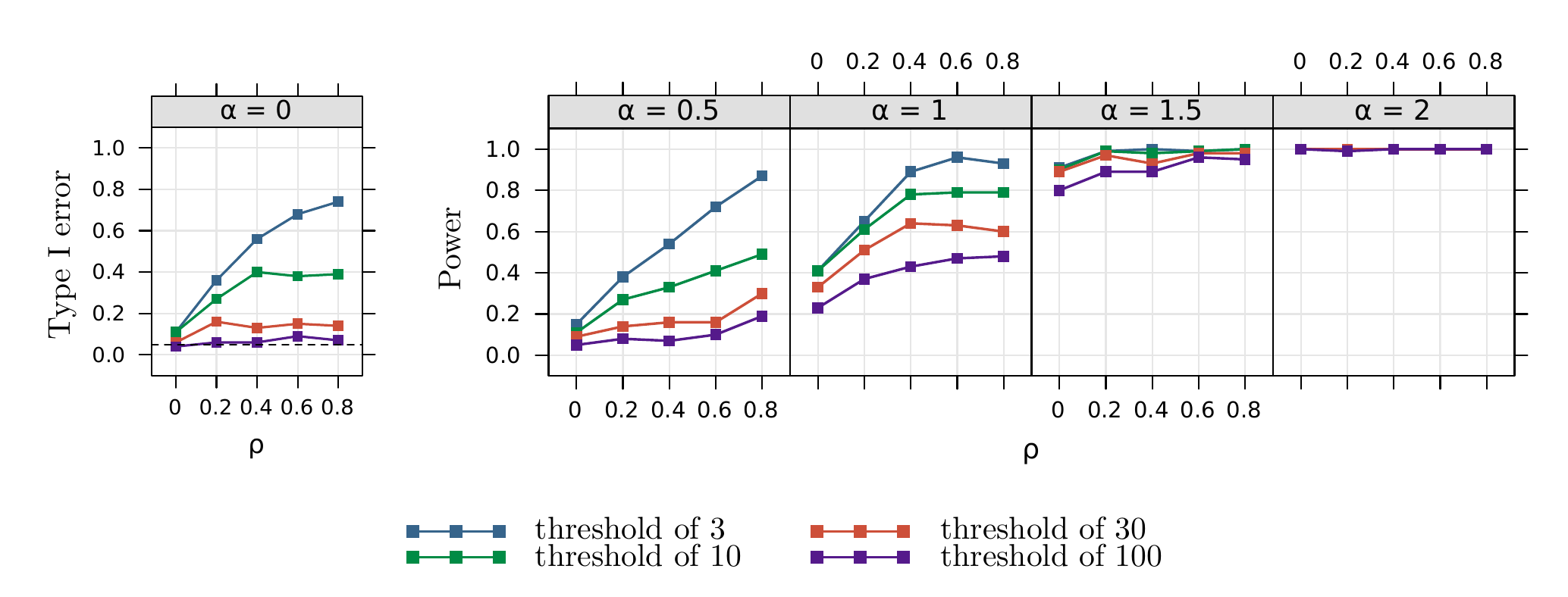}
    \caption{Simulation study: Type I error and power curves depending on the chosen threshold for the Bayes Factor to select the frailty values under $\mathcal{H}_0$ or under $\mathcal{H}_1^{(w^*)}$. $\alpha$ is the parameter that controls the cluster intensity and $\rho$ controls the spatial correlation.}
    \label{fig:perf_car_thresholds}
\end{figure}

Figure \ref{fig:perf_car_thresholds} shows the type I error and the power curves of the proposed method with several thresholds for the Bayes Factor (3, 10, 30 and 100). The thresholds of 3 and 10 do not maintain a stable type I error according to the values of $\rho$. The threshold of 100 maintains the type I error very well but is very conservative. The threshold of 30 seems to be a good compromise.  \\

\begin{figure}[h!]
\begin{minipage}{\linewidth}
(a)\\
\includegraphics[width=\textwidth]{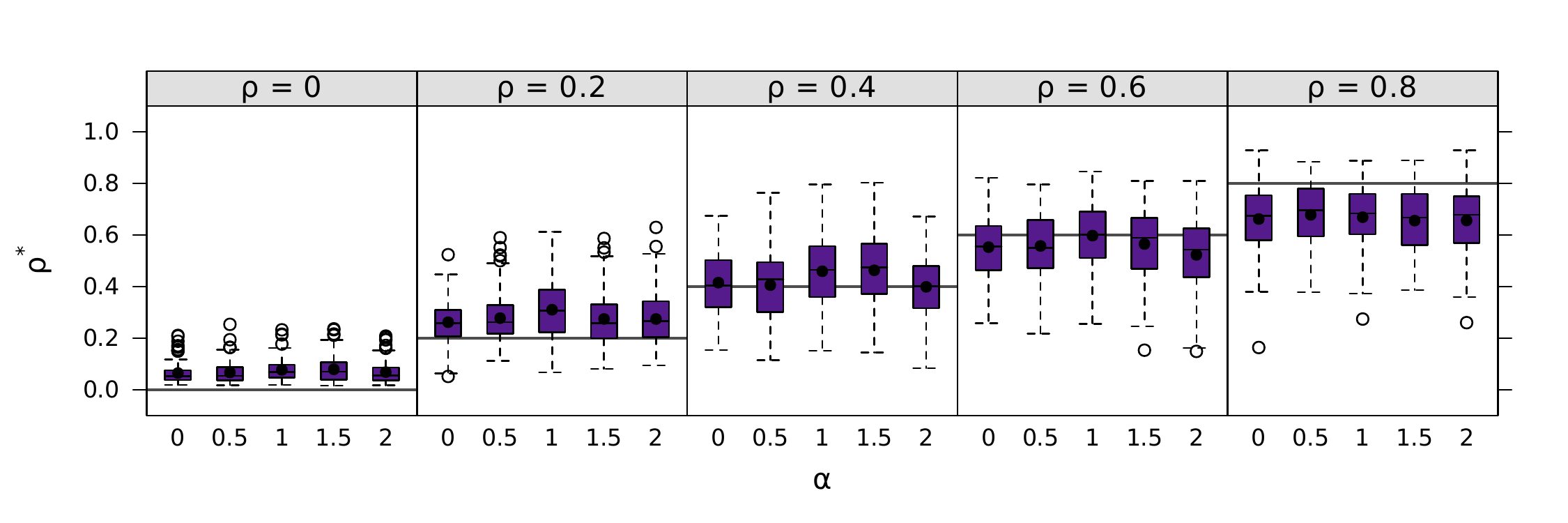}
\end{minipage}
\begin{minipage}{\linewidth}
(b) \\
\includegraphics[width=\textwidth]{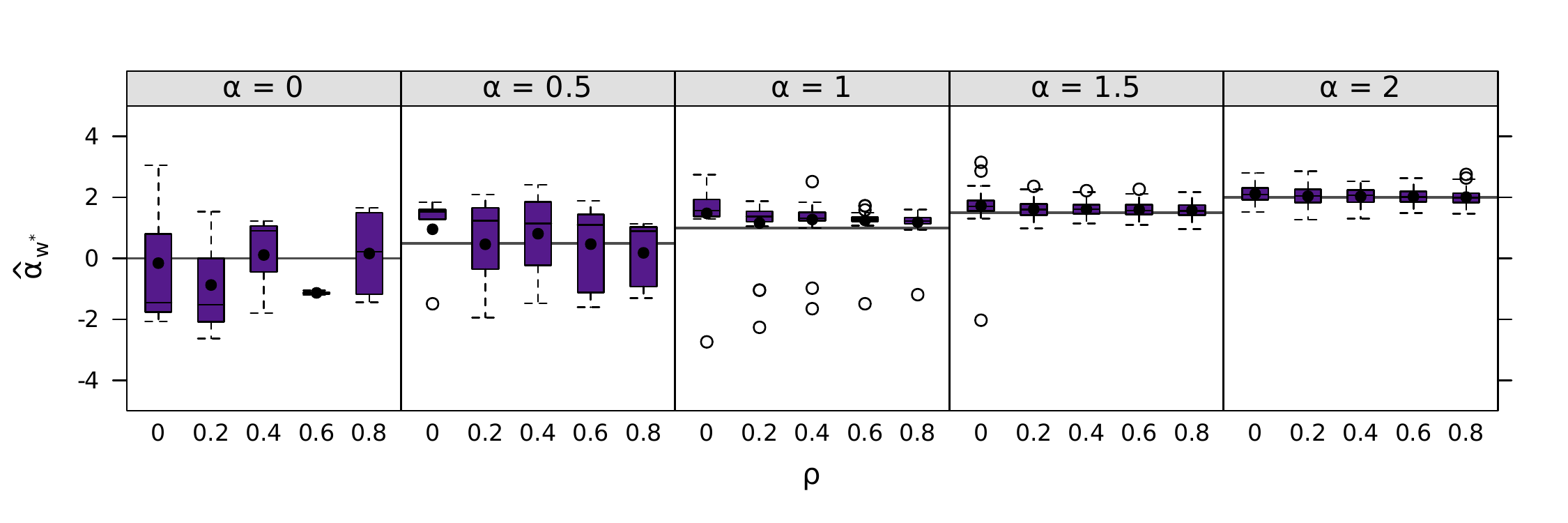}
\end{minipage}
\caption{Simulation study: the selected $\rho^*$ according to the parameters $\rho$ and $\alpha$ (panel (a)) and $\hat{\alpha}$ obtained with INLA when we select $\mathcal{H}_1$ according to the Bayes Factor criterion (panel (b)) with a threshold of 100. The main horizontal lines correspond to the true value of the parameters $\rho$ and $\alpha$ in panels (a) and (b) respectively and the black points represent the mean estimates obtained.}
\label{fig:boxplot_rhos_alphascar_100}
\end{figure}

However, we decided to investigate whether the estimates of the parameters with a threshold of 100 were better (less biased) than with a threshold of 30. Figure \ref{fig:boxplot_rhos_alphascar_100} shows that this is not the case.

\subsection{Type I error in presence of censoring} \label{appendix:typeIcensure}

\begin{figure}[h!]
    \centering
    \includegraphics[width=0.9\textwidth]{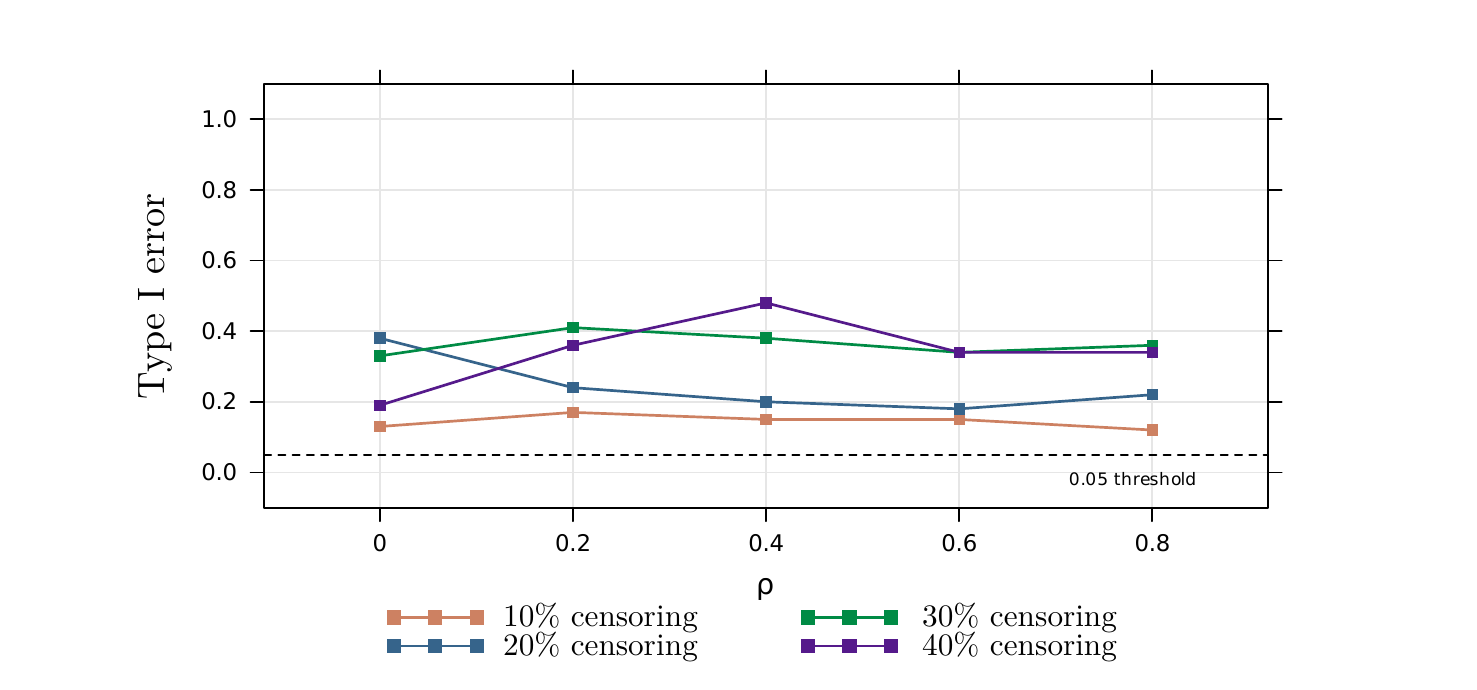}
    \caption{Simulation study: Comparison of the type I errors in the presence of censoring. $\rho$ controls the spatial correlation.}
    \label{fig:typeIcensure}
\end{figure}

Figure \ref{fig:typeIcensure} shows that the type I error increases when the proportion of censoring increases.

\section{Supplementary Materials of the application} \label{appendix:application}

Table \ref{tab:confounders} describes the confounding factors for the detection of clusters of abnormal survival times after starting dialysis in people aged 70 and over in the Nord-Pas-de-Calais region.

\begin{table}[h!]
\centering
\caption{Description of the confounding factors for the detection of clusters of abnormal survival times in ESRD elderly patients in northern France, 2004-2020.}
\begin{tabular}{lc}
\hline
\textbf{Variable} & \textbf{Overall}, $N = 6071^{1}$ \\
\hline
Age (in years) & 79.0 (74.8;83.5) \\
\hline
Body mass index (in $\text{kg/m}^2$) & 26.2 (23.0;30.1) \\
\hline
Sex (Woman) & 2582/6071 (42.5\%) \\
\hline
Type of nephropathy & \\
\hspace{0.7cm} Polycystic kidney disease & 104/5342 (1.9\%) \\
\hspace{0.7cm} Primitive glomerulonephritis & 427/5342 (8.0\%) \\
\hspace{0.7cm} Hypertension or vascular & 2048/5342 (38.3\%) \\
\hspace{0.7cm} Diabetic nephropathy & 1683/5342 (31.5\%) \\
\hspace{0.7cm} Pyelonephritis & 283/5342 (5.3\%) \\
\hspace{0.7cm} Other & 797/5342 (14.9\%) \\
\hline
Number of cardiovascular comorbidities & \\
\hspace{0.7cm} None & 962/5380 (17.9\%) \\
\hspace{0.7cm} One & 1566/5380 (29.1\%) \\
\hspace{0.7cm} At least two & 2852/5380 (53.0\%) \\
\hline
Diabete (Yes) & 3059/5960 (51.3\%) \\
\hline
Chronic respiratory disease (Yes) & 1062/5793 (18.3\%) \\
\hline
Respiratory assistance (Yes) &  380/5770 (6.6\%) \\
\hline
Cirrhosis (Yes) &  148/5804 (2.5\%) \\
\hline
Severe behavioral disorder (Yes) &  257/5532 (4.6\%) \\
\hline
Mobility & \\
\hspace{0.7cm} Autonomous walking &  3262/4930 (66.2\%) \\
\hspace{0.7cm} Need for a third party &  1205/4930 (24.4\%) \\
\hspace{0.7cm} Total disability &  463/4930 (9.4\%) \\
\hline
Hemoglobin level ($<11\text{g/dL}$) &  3760/5479 (68.6\%) \\
\hline
Albumin level ($<35\text{g/dL}$) &  2525/4786 (52.8\%) \\
\hline
Dialysis method &  \\
\hspace{0.7cm} Haemodialysis &  5330/6071 (87.8\%) \\
\hspace{0.7cm} Peritoneal dialysis &  741/6071 (12.2\%) \\
\hline
Emergency start (Yes) &  1776/5470 (32.5\%) \\
\hline
Active malignancy (Yes) &  586/5819 (10.1\%) \\ 
\hline
Glomerular filtration rate & \\
\hspace{0.7cm} $<7\text{mL/min/1.73m}^2$ &  882/5285 (16.7\%) \\
\hspace{0.7cm} $7-10\text{mL/min/1.73m}^2$ &  1535/5285 (29.0\%) \\
\hspace{0.7cm} $>10\text{mL/min/1.73m}^2$ &  2868/5285 (54.3\%) \\
\hline
Year of treatment initiation & \\
\hspace{0.7cm} 2004-2009 &  1928/6071 (31.8\%) \\
\hspace{0.7cm} 2010-2015 &  2258/6071 (37.2\%) \\
\hspace{0.7cm} 2016-2020 &  1885/6071 (31.0\%) \\
\hline
\end{tabular}
\flushleft{$^1$ Number of observed/ Total number of observed (\%) for qualitative variables \\
\hspace{0.17cm} Median (Interquartile range) for quantitative variables}
\label{tab:confounders}
\end{table}

\begin{figure}[h!]
\centering
\begin{minipage}{0.49\linewidth}
(a) \\
\includegraphics[width=0.9\textwidth]{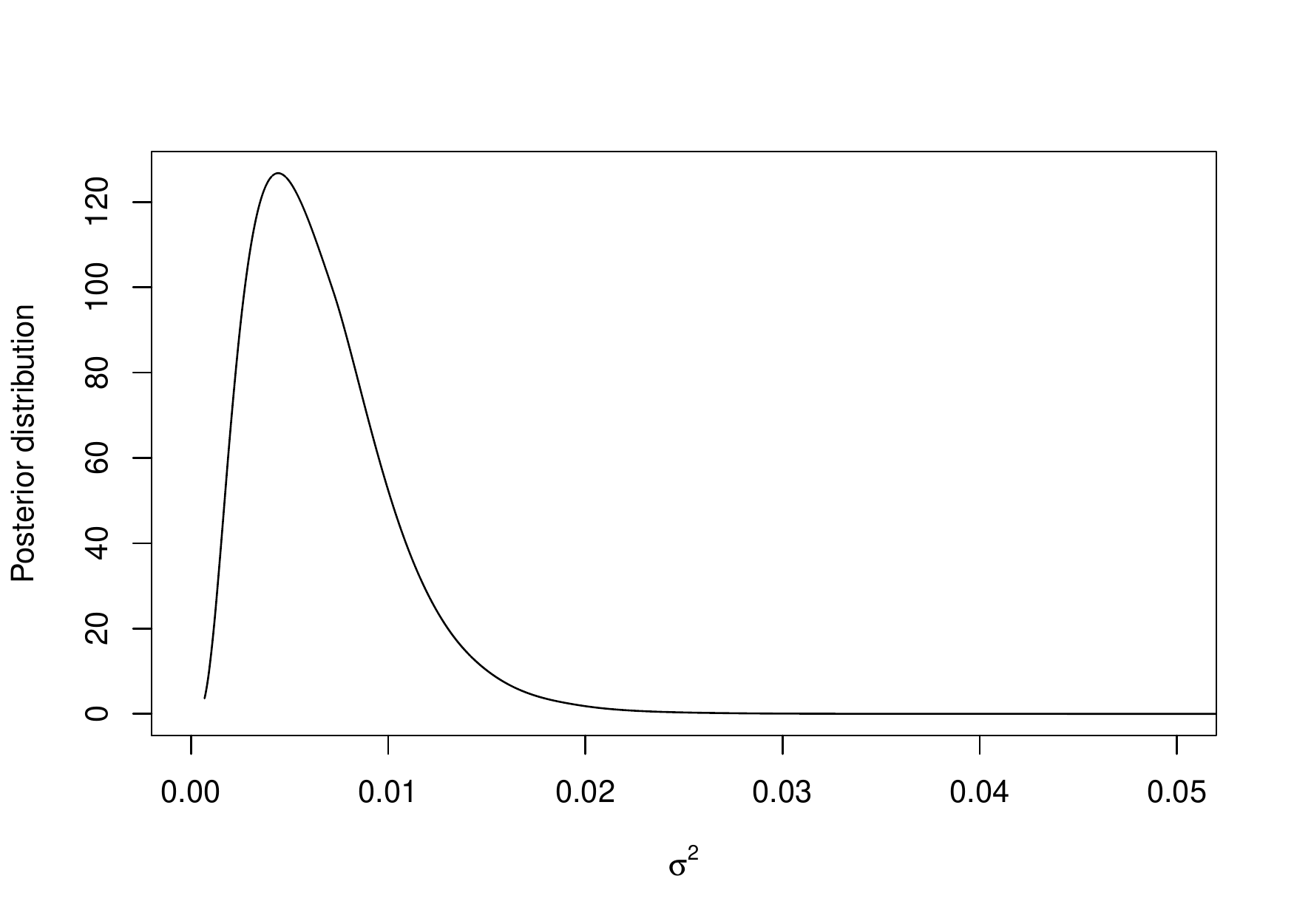}
\end{minipage}
\begin{minipage}{0.49\linewidth}
(b) \\
\includegraphics[width=0.9\textwidth]{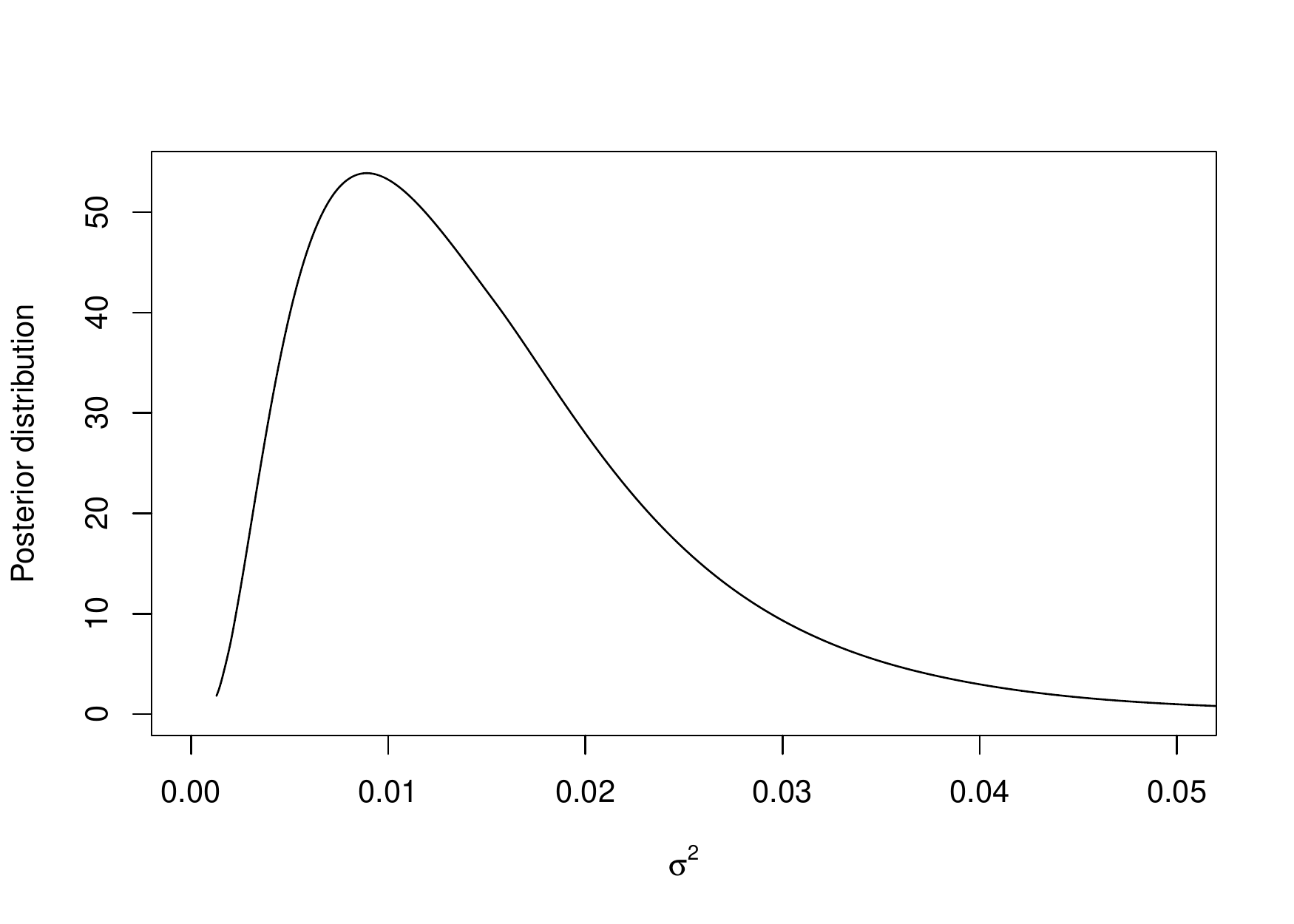}
\end{minipage}
\begin{minipage}{0.49\linewidth}
(c) \\
\includegraphics[width=0.9\textwidth]{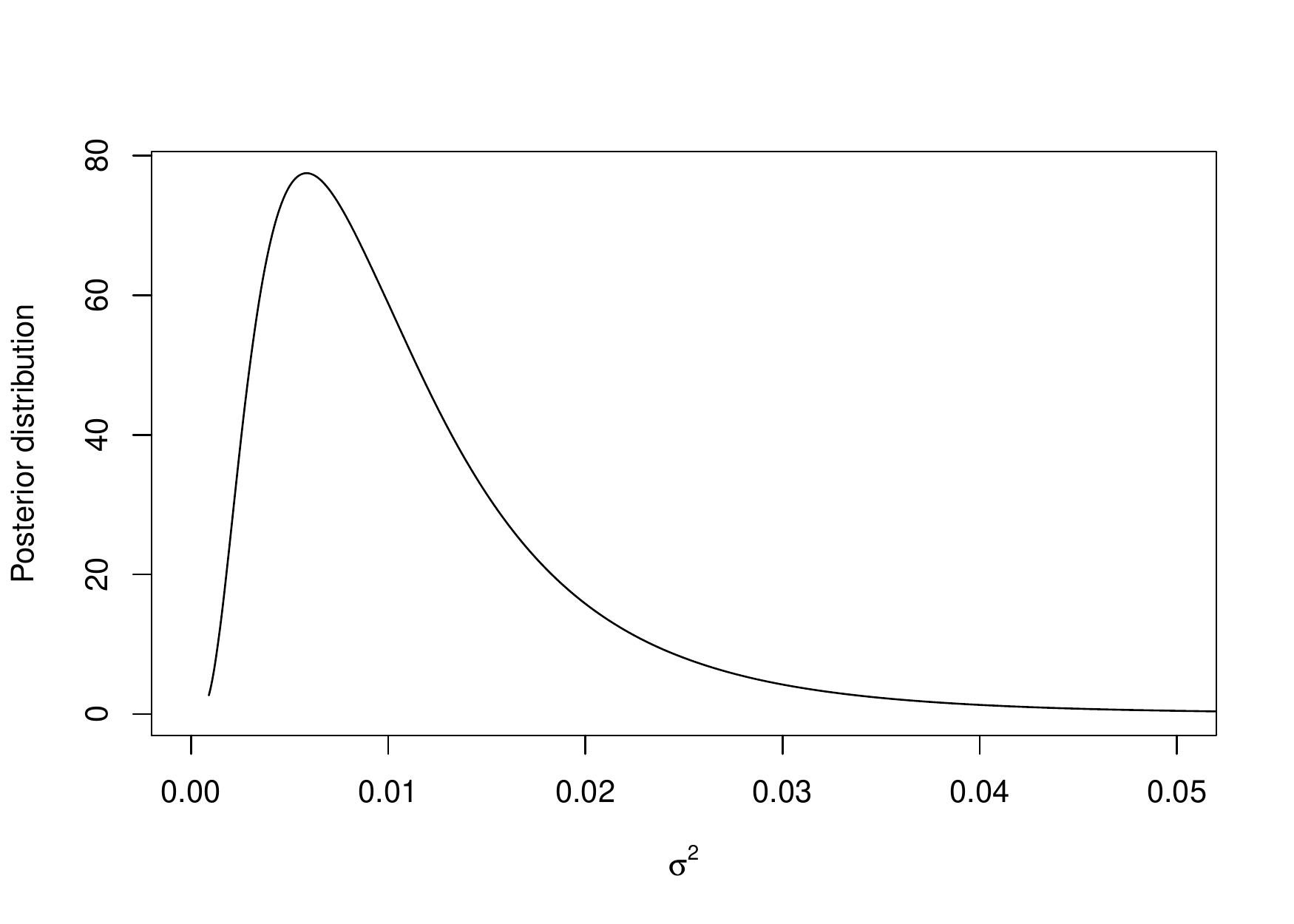}
\end{minipage}
\begin{minipage}{0.49\linewidth}
(d) \\
\includegraphics[width=0.9\textwidth]{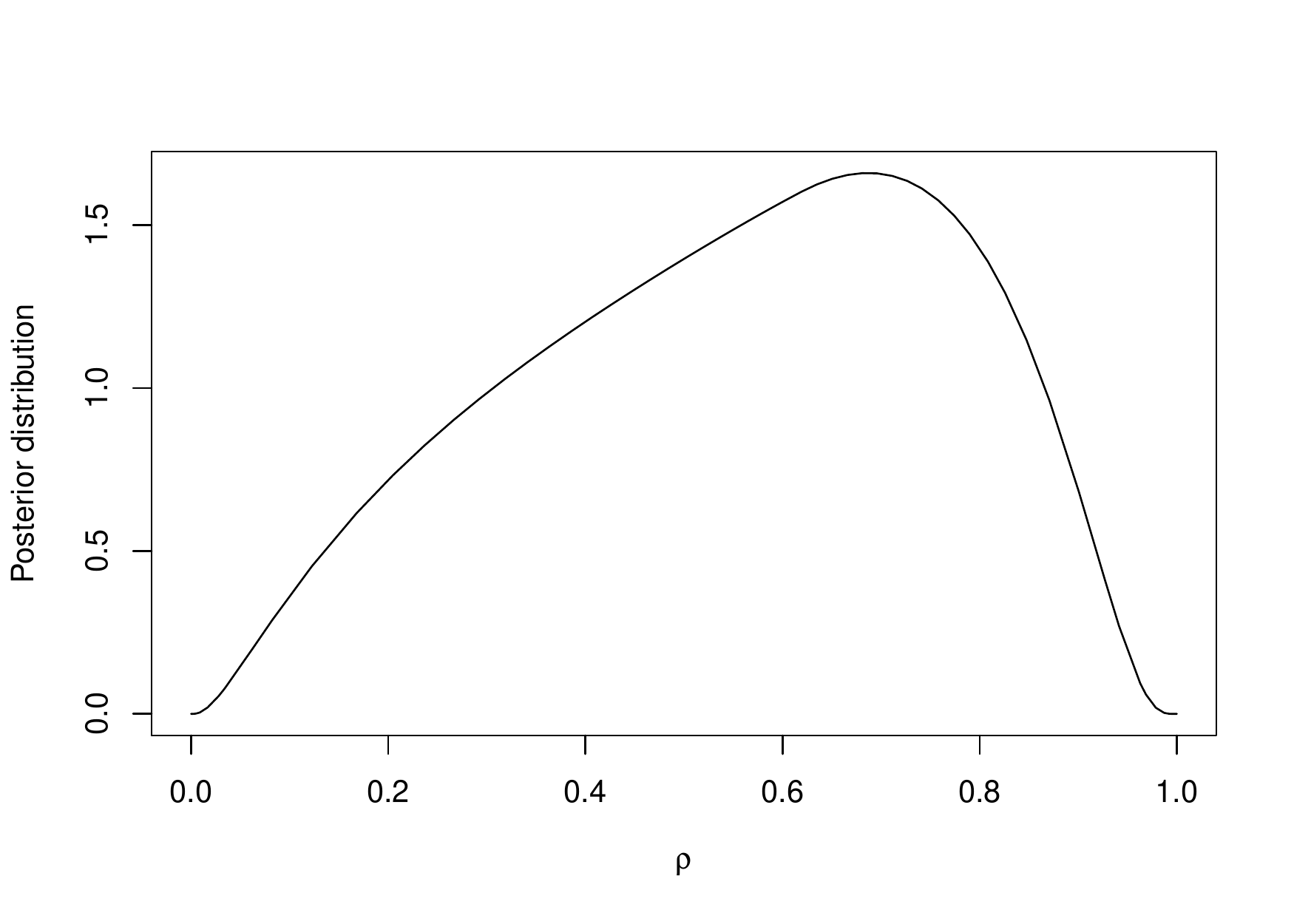}
\end{minipage}
\caption{Posterior distribution of the frailty variance with the i.i.d. (panel (a)), CAR (panel (b)) and ICAR (panel (c)) models and posterior distribution of the spatial correlation parameter $\rho$ obtained with the CAR model (panel (d))}
\label{fig:posteriors}
\end{figure}

Figure \ref{fig:posteriors} shows the posterior distributions of the frailty variance $\sigma^2$ with a i.i.d., a CAR and a ICAR model as well as the posterior distribution of the spatial correlation parameter $\rho$ of the frailty CAR model.

\begin{figure}[h!]
\centering
\begin{minipage}{0.49\linewidth}
(a) \\
\includegraphics[width=0.9\textwidth]{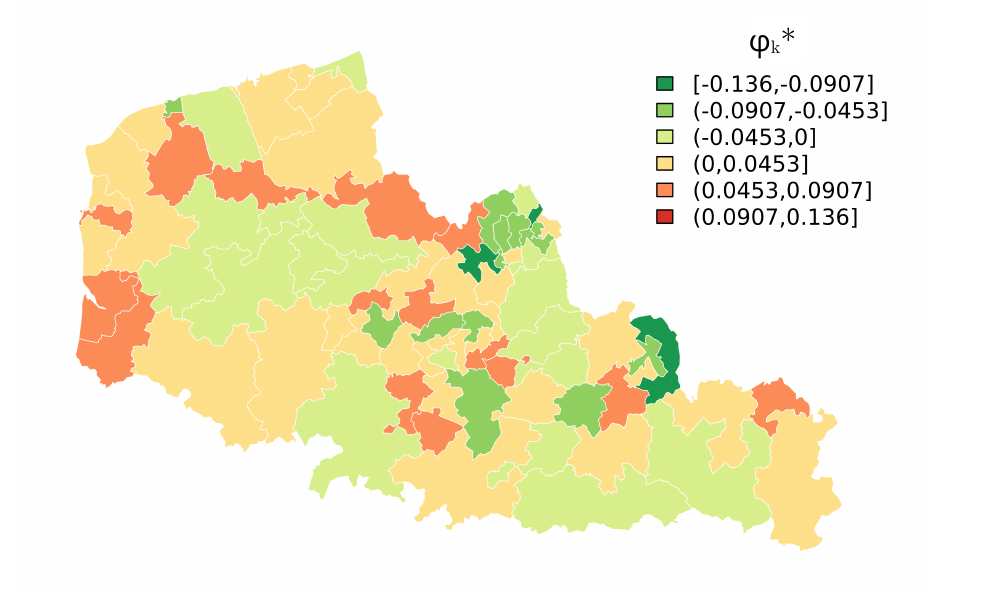}
\end{minipage}
\begin{minipage}{0.49\linewidth}
(b) \\
\includegraphics[width=0.9\textwidth]{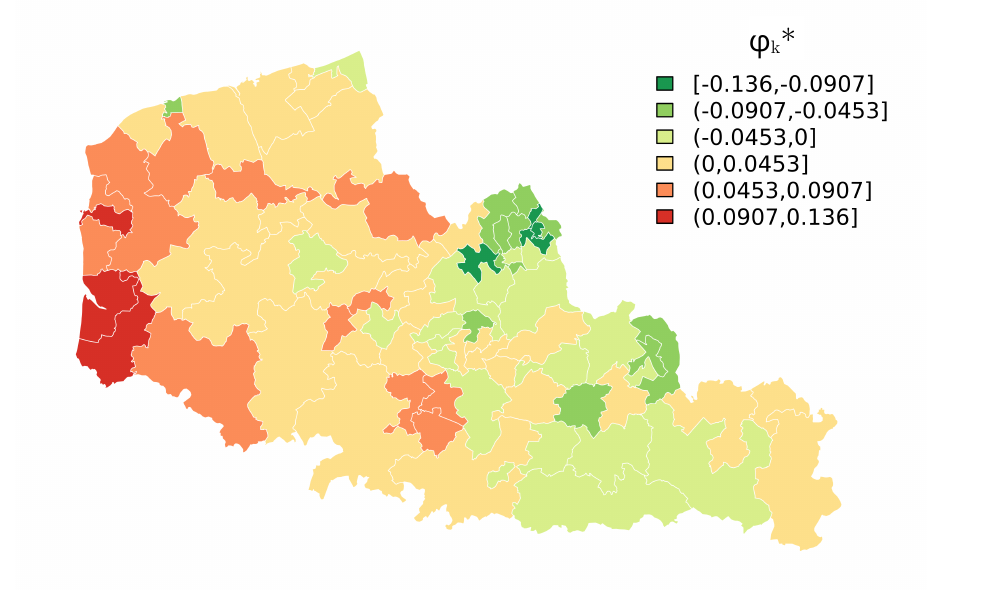}
\end{minipage}
\begin{minipage}{0.49\linewidth}
(c) \\
\includegraphics[width=0.9\textwidth]{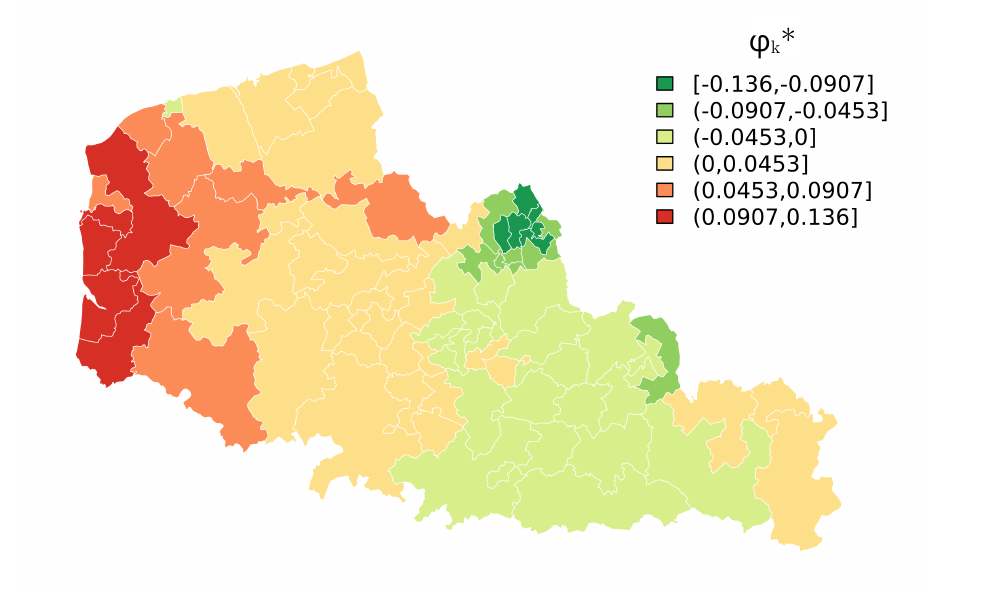}
\end{minipage}
\caption{Estimated frailties $\varphi_k^*$ with the i.i.d. (panel (a)), CAR (panel (b)) and ICAR (panel (c)) models}
\label{fig:frailties}
\end{figure}

Figure \ref{fig:frailties} presents the estimated $\varphi_k$ for each model: the i.i.d. model, the Leroux CAR model and the ICAR model.

\end{document}